\documentclass[%
 twocolumn,
superscriptaddress,
 amsmath,amssymb,
prx,
floatfix,
]{revtex4-2}

\usepackage{graphicx}
\usepackage{dcolumn}
\usepackage{bm}

\usepackage{tablefootnote}
\usepackage{threeparttable}

\usepackage{xcolor}
\usepackage[final]{changes}
\definechangesauthor[color=blue]{common}
\definechangesauthor[color=red, name={Paulo Santos}]{PVS}
\definechangesauthor[color=green, name={Alexander Kuznetsov}]{AK}
\setauthormarkup{}

\newcommand{\rPVS}[2]{\replaced[id=PVS]{#1}{#2}}
\newcommand{\aPVS}[1]{\added[id=PVS]{#1}}
\newcommand{\dPVS}[1]{\deleted[id=PVS]{#1}}
\newcommand{\cPVS}[1]{\comment[id=PVS]{#1}}

\definecolor{AK_color}{rgb}{1,0.25,0.1}
\definecolor{AK_HighlightColor}{rgb}{1,0.25,0.5}

\newcommand{\dAK}[1]{\deleted[id=AK]{#1}}			

\newcommand{\lSAW}{{\lambda_{\mathrm{SAW}}}}		
\newcommand{\kSAW}{k_\mathrm{SAW}}			
\newcommand{\fSAW}{f_\mathrm{SAW}}			
\newcommand{\TSAW}{T_\mathrm{SAW}}			
\newcommand{\pSAW}{\phi_\mathrm{SAW}}			

\usepackage{float}
\usepackage{upgreek}
\usepackage{makeidx}
\DeclareGraphicsExtensions{.png,.jpg,.pdf,.mps,.gif,.bmp,.eps}
\graphicspath{{figures/}}
\usepackage{physics}

\renewcommand\zeta{\xi}

\begin{document}

\preprint{SAW-OPO}

\title{Dynamically Tuned Arrays of Polariton Parametric Oscillators}


\author{Alexander S. Kuznetsov}
\email{kuznetsov@pdi-berlin.de}
\affiliation{Paul-Drude-Institut f{\"u}r Festk{\"o}rperelektronik, Leibniz-Institut im Forschungsverbund Berlin e.~V., Hausvogteiplatz 5-7, 10117 Berlin, Germany }

\author{Galbadrakh Dagvadorj}
\affiliation{Department of Physics and Astronomy, University College London, Gower Street London WC1E 6BT, United Kingdom }
\affiliation{Department of Physics, University of Warwick, Coventry CV4 7AL, United Kingdom}

\author{Klaus Biermann}
\affiliation{Paul-Drude-Institut f{\"u}r Festk{\"o}rperelektronik, Leibniz-Institut im Forschungsverbund Berlin e.~V., Hausvogteiplatz 5-7, 10117 Berlin, Germany }

\author{Marzena Szymanska}
\affiliation{Department of Physics and Astronomy, University College London, Gower Street London WC1E 6BT, United Kingdom }

\author{Paulo V. Santos}
\email{santos@pdi-berlin.de}
\affiliation{Paul-Drude-Institut f{\"u}r Festk{\"o}rperelektronik, Leibniz-Institut im Forschungsverbund Berlin e.~V., Hausvogteiplatz 5-7, 10117 Berlin, Germany }

\date{\today}

\begin{abstract}
Optical parametric oscillations (OPOs) -- a non-linear process \rPVS{involving}{of} the coherent coupling of an optically excited two particle pump state to a signal and an idler states with different energies --  is a relevant mechanism for optical amplification as well as for the generation of correlated photons. OPOs require states with well-defined  symmetries and energies: the fine-tuning of material properties and  structural dimensions to create these states remains a challenge for the realization of scalable OPO-based functionalities in semiconductor nanostructures. 
Here, we demonstrate a pathway towards this goal based on the control of confined microcavity exciton-polaritons modulated  by the  spatially and time varying dynamical potentials  produced  by a surface acoustic waves (SAW). The  exciton-polariton are confined in $\mu$m-sized intra-cavity traps fabricated by structuring a planar semiconductor  microcavity during the epitaxial growth process. OPOs in these structures benefit from the enhanced non-linearities of confined systems. We show that SAW fields induce state-dependent and time-varying energy shifts, which enable the energy alignment of the confined levels with the appropriate symmetry for OPO triggering. Furthermore, the dynamic acoustic tuning, which is fully described by a theoretical model for the modulation of the confined polaritons by the acoustic field,  compensates for fluctuations in symmetry and dimensions of the confinement potential thus enabling a variety of dynamic  OPO regimes. The robustness of the acoustic tuning is demonstrated by the synchronous excitation of an array of confined OPOs using a single acoustic beam, thus opening the way for the realization of  scalable non-linear on-chip systems.
\end{abstract}

\maketitle



\section{\label{sec:Introduction} Introduction}


Microcavity exciton-polaritons (polaritons) are light-matter quasi-particles resulting from the strong coupling between photons confined in a semiconductor microcavity (MC) with excitons in a quantum well (QW) embedded in the MC spacer \cite{Weisbuch92a}. Polaritons inherit the low effective mass from their photonic component, which gives them spatial coherence lengths of several $\mu$m. The polariton properties can thus be modified by confinement within $\mu$m dimensions as compared to the nm dimensions normally required to induce quantum shifts in electronic systems. In addition, the excitonic  component of polaritons gives rise to inter-polariton interactions and, thus, non-linearities much stronger than between photons. Finally, polaritons are composite bosons and can form Bose-Einstein-like condensates (BECs) at temperatures several orders of magnitude higher than cold atoms \cite{Kasprzak_N443_409_06}.

The mixed light-matter nature of polaritons brings the rich physics of correlated systems to an all-semiconductor platform \cite{Carusotto_RMP85_299_13,Sanvitto_NM15_1061_16}. It was early recognized that the strong inter-polariton interactions and the peculiar shape of the polariton energy dispersion enable stimulated parametric \rPVS{amplification}{scattering} with very large gain \cite{Savvidis_PRL84_1547_00,Saba01a} as well as  optical parametric oscillations (OPOs) \cite{,Baumberg00a,Saba01a,Langbein_PRB70_205301_04}. 
Here, two pump ($p$) polaritons resonantly excited at the inflection point of the lower polariton dispersion can scatter into  signal ($s$) and idler ($i$) states while conserving energy and momentum \cite{Ciuti_PRB69_245304_04,Savasta_PRL94_246401_05,Romanelli_PRL98_106401_07,Portolan_JPCS210_12033_10}.  
\rPVS{The }{In addition, the} OPO process is  a convenient approach for direct excitation of polariton condensates by stimulated scattering to the $s$ states at the bottom of the dispersion \cite{Baumberg00a}. This excitation scheme avoids the formation of a high-density reservoir of excitons with high in-plane momentum, which normally occurs for non-resonant optical injection.

\rPVS{The OPO process provides a pathway for the efficient generation of correlated and entangled photons. Different approaches have been proposed to enhance the efficiency of the process including }{ Approaches to enable OPO scattering  include} the engineering of the polariton density of states by using multiple cavities \cite{Diederichs_N440_904_06} and spatial confinement \cite{Ferrier_APL97_31105_10}. Confinement creates a discrete spectrum of polariton levels. The latter can act as pump, signal, and idler states, provided that the symmetry and energy spacing required for OPO  are satisfied. This approach profits from the high density of polaritons that can be excited in confined potentials, which enhances the non-linear interactions required for OPO formation. In addition, since OPO properties are controlled by the dimensions of $\mu$m-sized polariton traps,  a further advantage of confinement is the scalability arising from the combination of multiple OPO structures on the same polariton MC. 
The design of confined OPO levels with  equally spaced pump, idler, and signal states with the  appropriate symmetries to enable mutual non-linear interactions remains, however,  a challenging task.   \rPVS{The studies in Ref.~}{Work~}\cite{Ferrier_APL97_31105_10} show that this requirement  can be satisfied by the confined states of square pillars etched in a (Al,Ga)As polariton MC. However, even a smallest deviation in the potential shape results in non-equidistant energy spectrum, thus preventing OPO excitation. 

In this work, we demonstrate a pathway for the efficient generation of confined polariton states with the appropriate symmetry and energy alignment for OPOs via the dynamic energy tuning by a surface acoustic wave (SAW). The studies are carried out in intra-cavity polariton traps defined in the spacer layer of an (Al,Ga)As MC fabricated by molecular beam epitaxy (MBE), cf. Fig.~\ref{FigStr}. 
We show that the spatially dependent SAW fields induce state-dependent energy shifts of the confined levels, which enable the  alignment of levels with the appropriate symmetry for OPO triggering. Spatially resolved wave function maps of OPO states prove that the signal and idler states must have same parity, in agreement with the predictions of a model developed to account for non-linear interactions between confined levels in the contact approximation regime~\cite{Carusotto_RMP85_299_13}. Time-resolved investigations prove the dynamic character of the acoustic OPO triggering at multiples of the SAW frequency. Finally, the acoustic modulation \rPVS{enables robust}{provides a robust approach for} OPO triggering over a wide  range of excitation conditions. In particular, we demonstrate the synchronous tuning of an array of confined OPOs using a single SAW beam, thus proving the feasibility of scalable OPO systems. 

\section{\label{Experimental_Details}Experimental Details}

\begin{figure}[thbp]
\includegraphics[width=1\columnwidth, keepaspectratio=true]{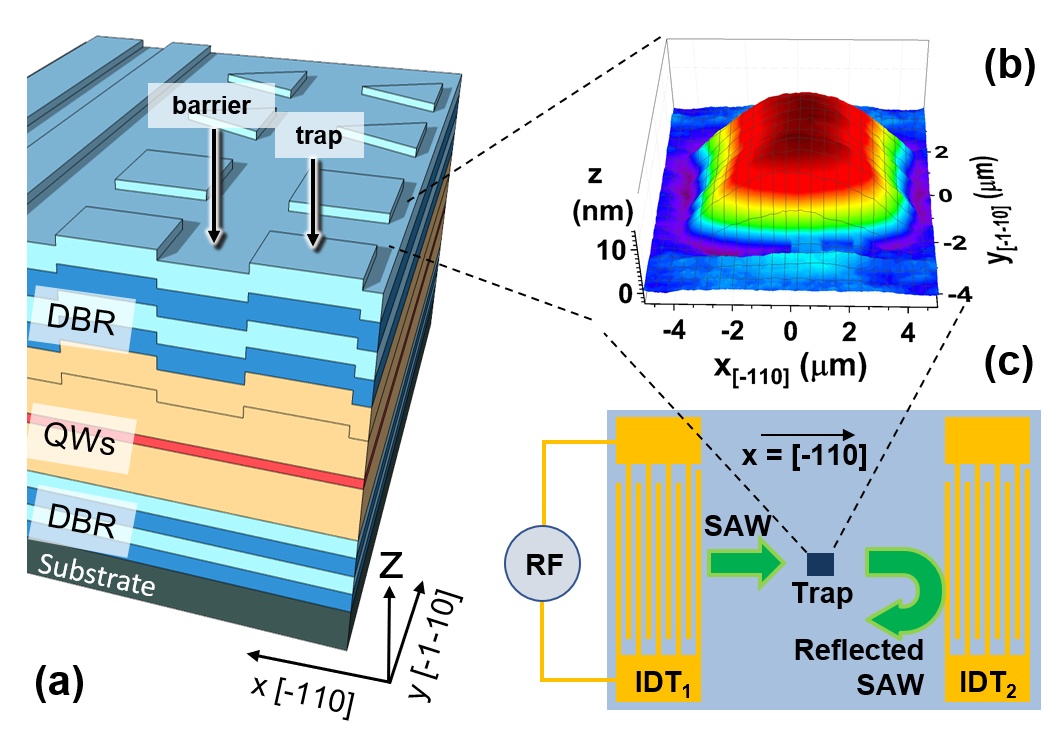}
    \caption{{\bf Intra-cavity polariton traps.} (a) Polariton microcavity with intra-cavity  traps defined by structuring the spacer region during the growth process.  (b) Atomic force micrograph showing the surface relief of a nominally $4\times4~\mu$m$^2$ trap. (c) Surface acoustic wave (SAW) resonator containing an intra-cavity polariton trap.
}
    \label{FigStr}
\end{figure}

Confined polariton states employed here originate from $\mu$m-sized intra-cavity traps created within the spacer layer of an (Al,Ga)As  MC. The traps were produced by etching the spacer layer of the MC between growth steps by molecular beam epitaxy (MBE). The sample was grown on a GaAs (001) substrate having the structure schematically illustrated in Fig.~\ref{FigStr}(a). During the MBE growth run, the lower distributed Bragg reflector (DBR) and the MC spacer region containing three pairs of 15 nm-thick GaAs QWs centered at the anti-nodes of the optical field were deposited and then terminated by a 120~nm-thick  Al$_{0.15}$Ga$_{0.85}$As layer. The sample was then removed from the MBE chamber  and patterned by means of photolithography and wet chemical etching to form 12 nm-high and a few $\mu$m-wide mesas with different shapes. For the final growth step, the sample was reinserted in the MBE chamber for the deposition of the upper DBR. The lower and upper DBRs consist of \aPVS{58.7~nm and 65.8~nm-thick} pairs of Al$_{x_1}$Ga$_{1-x_1}$As/Al$_{x_2}$Ga$_{1-x_2}$As with different Al compositions $x_1=0.15$ and $x_2=0.75$

The etching of the MC spacer layer results in a blueshift of the bare optical mode of the MC by 9~meV. As a consequence,  the energy of the lower polariton mode in the etched regions becomes  approx. $E_b=5.5$~meV higher than in the non-etched areas. Polariton traps with an energy barrier $E_b$  can then be formed by enclosing a $\mu$m-sized non-etched region by etched areas. Due to the conformal nature of the MBE growth, the lateral dimensions of the traps can be estimated by measuring the surface relief of the MC, as illustrated in the Fig.~\ref{FigStr}(b) for a square trap with nominal lateral dimensions of $4\times4~\mu$m$^2$. The anisotropic MBE growth yields traps with different profiles along the trap sides oriented along the $x=\left[\bar110\right]$ and $y=\left[\bar1\bar10\right]$ surface directions. A detailed analysis of AFM profiles from the traps presented in Secs.~\ref{SM:Intra-cavity_polariton_traps} of the  Supplementary Material (SM) shows that the confinement potential is mirror-symmetric with respect to vertical planes $x=0$ and $y=0$ [cf. Fig.~\ref{FigStr}(b)] but with different profiles along the two directions. 

The spectroscopic studies were performed at a temperature between 6 and 10~K in an optical cryostat with radio-frequency (rf) feedthroughs for the excitation of SAWs. We have measured the photoluminescence (PL) of  small square traps (dimensions $\le 4~\mu$m) placed within a SAW  delay line formed by two single-finger interdigital transducers (IDTs). The latter were designed for launching SAWs along the $x||[\bar110]$ surface direction with a wavelength of $8~\mu$m (corresponding to an acoustic frequency $\fSAW=383.66$~MHz at 10~K). The delay line forms an acoustic resonator with a quality factor of 4700 \cite{PVS318}. Care was taken to match the location of the traps with the anti-nodes of the SAW strain field. Time-resolved studies of the OPO dynamics were carried out by detecting the PL with a streak camera synchronized with the rf-signal used to excite the SAW resonator.

\begin{figure*}[!thbp]
\includegraphics[width=0.95\textwidth, keepaspectratio=true]{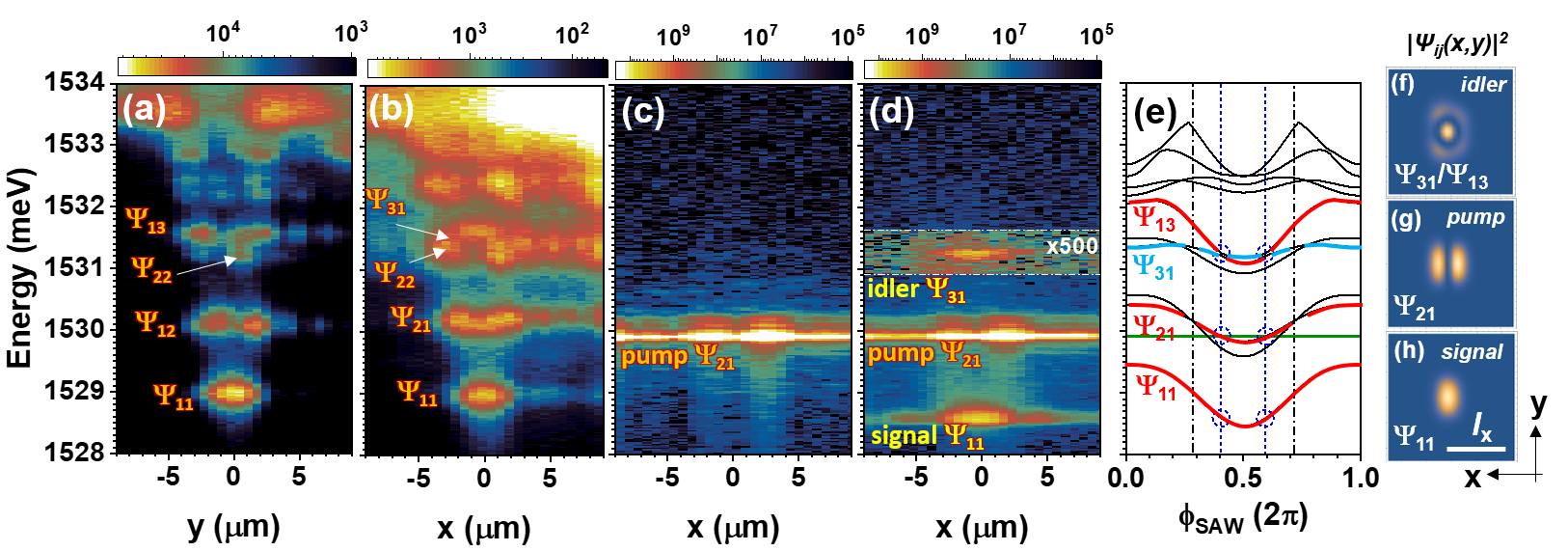}
    \caption{
{\bf OPO in a polariton trap with red-shifted pump}. Photoluminescence maps of the squared wave functions $|\Psi_\mathrm{ij}|^2$ ($i,j=1,2,3$) of confined polariton levels in a  $4\times4~\mu$m$^2$ intra-cavity trap projected on the  (a) $y=0$ and (b) $x=0$ planes  under the low-density, non-resonant optical excitation. 
(c)-(d) Corresponding spatially resolved PL maps recorded under quasi-resonant excitation of the  $\Psi_\mathrm{21}$ confined level in the (c) absence and (d) presence of a standing SAW along $x$ [cf.~Fig.~\ref{FigStr}(c)]. 
(e) Calculated energy evolution of the confined levels $\Psi_{m_x m_y}$  with the SAW phase, $\pSAW$: the labeled ones (thick red curves) are levels of the OPO set $O_2=\{\Psi_{11}, \Psi_{21} , \Psi_{31}\//\Psi_{13}\}$. \dAK{ with energies $E_{11}$, $E_{21}$, and $E_{31}$, respectively.} The dash-dotted black vertical lines designate SAW phase for which the levels correspond to the unmodulated ones. The green horizontal line marks the pump energy while the dashed vertical blue lines show the phases of equidistant separation between the levels within $O_2$. (f)-(h) Calculated  $|\Psi_\mathrm{m_x m_y}(x,y)|^2$ projections  on the $x$-$y$-plane corresponding to the $\phi_{SAW} = 0.415~(2\pi)$.
}
    \label{FigOPO}
\end{figure*}

\section{\label{Results}Results}

\subsection{Confined polaritons in intra-cavity traps}
\label{Trapping_potential}

Figure~\ref{FigOPO}(a) and \ref{FigOPO}(b) compare PL spectral maps of a square trap with nominal dimensions of $4\times4~\mu$m$^2$ projected on the $x=0$ and $y=0$ planes, respectively. These maps were recorded  under low-density non-resonant optical excitation conditions by collecting the PL with spatial resolution along two perpendicular directions, thus yielding the projection of the squared wave functions $|\Psi_\mathrm{ij}|^2$ ($i,j=1,2,\dots$) of the  confined polariton levels on the  $x=0$ and $y=0$ axes, respectively.  The $i$ and $j$ indices denote the number of lobes of $|\Psi_\mathrm{ij}|^2$ along the $x$ and $y$ directions, respectively. 

The polariton states in the intra-cavity traps can be approximated by those of a rectangular one with infinite barriers and  dimensions $\ell_x \sim\ell_y$ along the $x$ and $y$ directions, respectively, which can be classified by indices $(m_x, m_y)$, $m_i=1,2,\dots$ and $i=x, y$) according to: 

\begin{equation}
E_\mathrm{m_xm_y} =E_\mathrm{LP} + 
\frac{\hbar^2}{2m_p} \left[ \left(\frac{m_x-1}{l_x}\right)^2 + \left(\frac{m_y-1}{l_y}\right)^2  \right].
\label{EqE}
\end{equation}
 
\noindent Here $\hbar$ is Planck's constant,  $m_p$ the reduced polariton mass, and $E_\mathrm{LP}$ the lower polariton energy in the absence of lateral confinement. The corresponding wave functions can be written as:
\begin{eqnarray}
\Psi_\mathrm{m_xm_y}(x,y) =& \frac{\sqrt{(k_xk_y)}}{\pi} \times \\ \nonumber
&\left[ \right.
      \cos{\left(m_x k_x x + (m_x - 1) \frac{\pi}{2} \right)} \times  \\ \nonumber
     &\cos{\left(m_y k_y y + (m_y - 1) \frac{\pi}{2} \right)}
\left. \right],
\label{EqWF}
\end{eqnarray}
 
 \noindent with $k_i=\pi/\ell_i$.  \aPVS{Note that the wave function projections in  Figs.~\ref{FigOPO}(a) and \ref{FigOPO}(b) do not show modes with nodes on the $y$ or $x$ axis, respectively. As a consequence, some of the modes are either very weak or simply do not appear in the maps  (one example is the $\Psi_\mathrm{22}$ mode displayed in Fig.~\ref{FigSM1}(i)).}

\subsection{OPO in intra-cavity traps}
\label{OPO in intra-cavity traps}


 According to Eq.~(\ref{EqE}), the confined levels  with the set
 $\{\Psi_\mathrm{signal}, \Psi_\mathrm{pump} , \Psi_\mathrm{idler}\}=O_1=\{\Psi_\mathrm{11}, \Psi_\mathrm{12} \text{ or } \Psi_\mathrm{21}, \Psi_\mathrm{22}\}$ 
are equidistant in energy for $l_x=l_y$, thus satisfying the energy requirements for OPOs. The lowest three confined levels in Fig.~\ref{FigOPO}(a), however, do not follow this behavior. In addition, the levels  $\Psi_\mathrm{12}$ and $\Psi_\mathrm{21}$ are not degenerate, cf. Fig.~\ref{FigOPO}(a \& b). These discrepancies arise from the deviation of the confinement potential from a perfect square shape due to the anisotropic effects during the MBE overgrowth (cf. Sec.~\ref{SM:Intra-cavity_polariton_traps} for details) \cite{PVS312}. The impact of the anisotropic shape on the confined states can be reproduced by assuming  $l_x\neq\l_y$ in Eq.~(\ref{EqE}), which lifts the degeneracy of the $(m_x, m_y)$ and $(m_y, m_x)$ states with $(m_x\neq m_y)$ . 

The different inter-level spacings and the symmetry of the states (cf.~Sec.~\ref{Symmetry of the OPO states})  within  $O_1$ prevent OPO excitation by optically pumping of the $\Psi_\mathrm{21}$ level. An experimental implementation is illustrated in the PL map of Fig.~\ref{FigOPO}(c). Here, the trap was illuminated by a continuous wave (cw) laser beam \dPVS{(spot diameter of approx.  $50~\mu$m)} slightly red-shifted with respect to the $\Psi_\mathrm{21}$ emission in Fig.~\ref{FigOPO}(a)-(b), but within its linewidth. \aPVS{The angle of incidence of the laser   (of $10^\circ$) was chosen to match the emission peak in momentum space of the   $\Psi_\mathrm{21}$ state\dAK{(cf.~Fig.~\ref{FigOPO}(a)}. In addition, this configuration helps to avoid the specular reflection of the pump. Due to focusing, the excitation  beam has an angular spread of $\pm2^\circ$. The spatially resolved PL was collected for angles between $0^\circ$ and $\pm16^\circ$ with respect to the sample surface.} The map shows a slight increase of the Rayleigh scattering due to pump interactions with the lobes of the  $\Psi_\mathrm{21}$ state superimposed on a background of stray light from the pump laser. 

When the SAW is turned on, the intra-cavity trap becomes subjected to an effective modulation potential $V_\mathrm{SAW}$ given by \cite{PVS318}:
\begin{equation}
V_\mathrm{SAW}(x,t)=V_\mathrm{SAW,0} \cos{(\kSAW x)}\cos{(\pSAW)}.
\label{EqVSAW}
\end{equation}

\noindent Here, $\kSAW=2\pi/\lSAW$ and  $\pSAW=2\pi \fSAW t$ denote the wave vector and phase of the standing SAW field, respectively. Under the acoustic modulation, a signal-idler pair appears at energies equidistant to the pump energy, thus  signalizing  OPO triggering. A detailed analysis of the  dependence of the OPO excitation  on the frequency of the rf-drive applied to the IDT (cf.~Sec.~\ref{SM:Frequency}) reveals  that OPO states only appear for rf-frequencies matching the modes of the acoustic resonator, thus unambiguously proving that OPO triggering is induced by the SAW field.

The PL map of Fig.~\ref{FigOPO}(d) also yields information about the symmetry of the confined states participating in the OPO process. In fact, the pump state in this figure has two lobes along $x$, thus indicating that it corresponds to the $\Psi_{21}$ (rather than the closely lying $\Psi_{12}$ state). The idler state, in contrast, has a single lobe along $x$ located in-between the two lobes of the pump state. This emission pattern does not correspond to the one expected for the $\Psi_\mathrm{22}$ state but rather to a superposition of the state   $\Psi_{13}$ and $\Psi_{31}$. As will be shown in Sec.~\ref{Symmetry of the OPO states}, this state red-shifts under the  acoustic modulation to satisfy the OPO energy matching condition within the set of levels $O_2=\{\Psi_\mathrm{11}, \Psi_\mathrm{21}, \Psi_\mathrm{13}/ \Psi_\mathrm{31}\}$ displayed in the rightmost panels of Fig.~\ref{FigOPO}.

\begin{figure}[tbhp]
    \centering
    \includegraphics[width=\columnwidth, angle=0, clip]{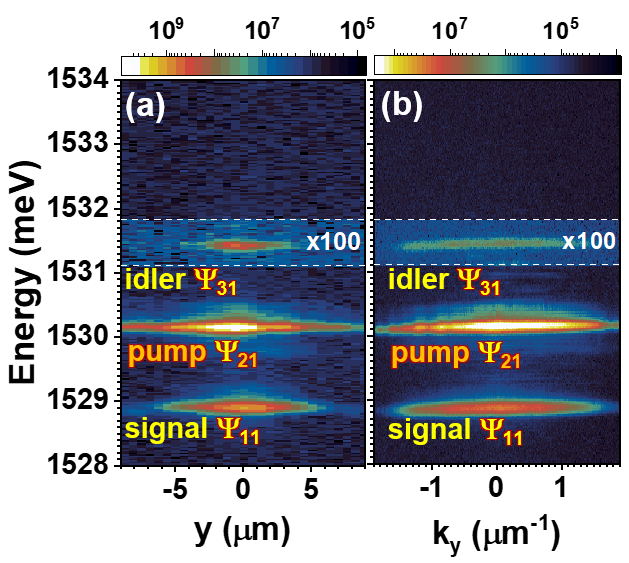}
    \caption{
{\bf OPO with resonant pump}. 
Acoustically driven OPO  detected by  (a) spatially and (b) momentum resolved PL. The PL maps were recorded along the direction $y$ perpendicular to the SAW propagation axis $x$. \dAK{The dashed horizontal line mark the energy of the $\Psi_\mathrm{21}$ state in Figs.~\ref{FigOPO}(a)-(v).} The optical pump is in resonance to the $\Psi_{21}$ level, cf. Fig. ~\ref{FigOPO}(a), and triggers a type $O_3=\{\Psi_{11}, \Psi_{21} , \Psi_{31}\}$ OPO.  
}
    \label{FigOPOb}
\end{figure}

Interestingly, the OPO level configuration is not unique and can be changed by varying the pump energy. Fig.~\ref{FigOPOb} displays PL maps recorded by setting the pump laser energy in resonance to the $\Psi_{21}$ level (in contrast to the red-shift pumping employed in Fig.~\ref{FigOPO}).  The real-space and momentum-space (angle-resolved) maps were recorded in the direction orthogonal to the SAW-propagation and optical excitation. While the OPO-pump state is still $\Psi_{21}$, the idler has single emission maximum  corresponding  to the $\Psi_{31}$ state. 
The excited OPO thus corresponds to  $O_3=\{\Psi_{11}, \Psi_{21}, \Psi_{31}\}$. We will show in Secs.~\ref{Symmetry of the OPO states} and \ref{Numerical simulations of the OPO dynamics} that this OPO mode set is in full agreement with a numerical model for the OPO based on the numerical solution of the Gross-Pitaevskii equation for this particular pump excitation energy.


\subsection{OPO switching dynamics}
\label{OPO_switching_dynamics}

\begin{figure}[tbhp]
    \centering
    \includegraphics[width=1.0\columnwidth, angle=0, clip]{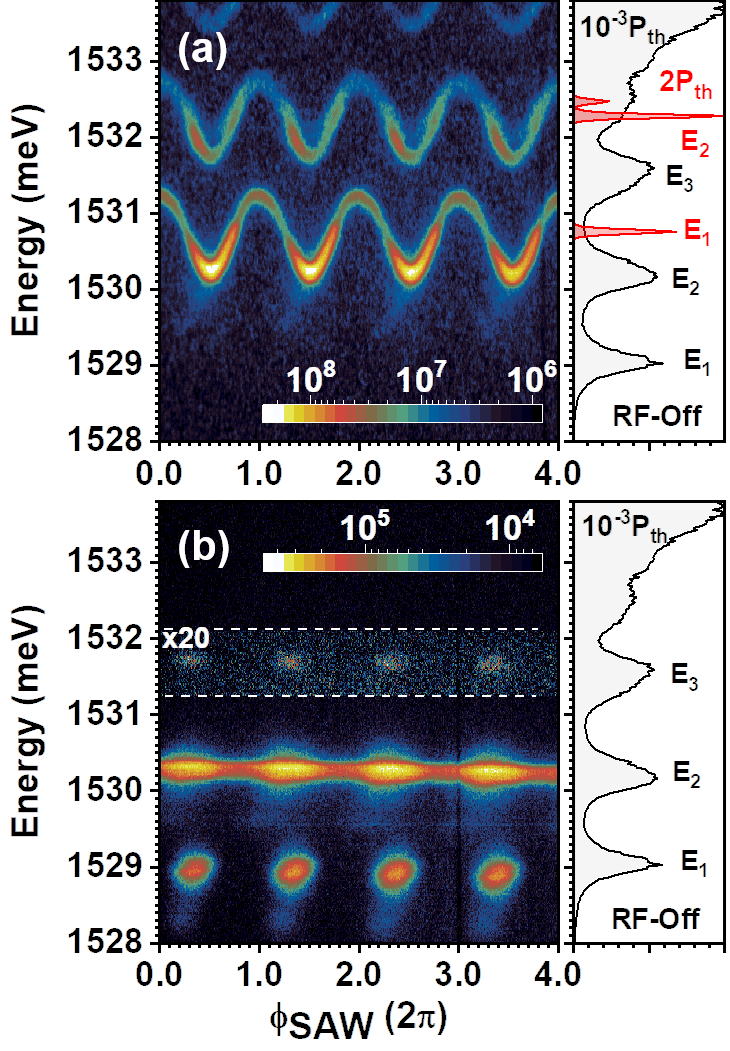}
    \caption{{\bf Time-resolved OPO triggering}. 
Time-resolved PL maps of the nominally $4\times4~\mu$m trap   recorded under a $\fSAW=383.69$~MHz ($\TSAW=2.6$ ns) SAW measured along y-axis perpendicular to the SAW direction. 
    (a) Map acquired under non-resonant optical excitation with a power density $2P_\mathrm{th}$, where $P_\mathrm{th}$ is the threshold power for the condensation. The right panel displays the spectral dependence of the time-integrated emission in the absence of a SAW for optical  excitation powers of $10^{-3} P_\mathrm{th}$ (black) and $2P_\mathrm{th}$ (red). 
    (b) OPO excitation by pumping the second confined level under the same acoustic excitation as in (a). \dAK{The ghost structures marked by an arrow are artifacts of the time-resolved setup.} The right panel displays the integrated PL in the absence of a SAW for optical  excitation powers of $10^{-3} P_\mathrm{th}$. The phases in (a) and (b) were not synchronized and were aligned manually using numerical simulations of Sec.~\ref{SM:Intra-cavity_polariton_traps}.
}
    \label{FigTOPO}
\end{figure}

The dynamic character of the acoustic modulation was investigated by analyzing the time dependent PL from the  $4\times4~\mu$m intracavity trap using a streak camera. Figure~\ref{FigTOPO}(a) displays the PL response of polariton condensates in the first and second confined levels recorded under a SAW. This spectrum was acquired under non-resonant excitation at 1.54 eV with an optical power equal to twice the condensation threshold ($P_\mathrm{th}$) \cite{Kasprzak_N443_409_06}. The right panel compares the time-integrated PL spectra recorded in the absence of a SAW under weak optical excitation ($10^{-3} P_\mathrm{th}$) and in the condensation regime ($2P_\mathrm{th}$). The large energy blue-shift of the condensate energies with respect to the ones measured at low excitation are attributed to the polariton interactions with the excitonic reservoir produced by the non-resonant excitation. As discussed in detail in Ref.~\onlinecite{PVS318}, the modulation by the SAW field leads to a sinusoidal dependence of the emission energy of the confined states. 

The left panel in Fig.~\ref{FigTOPO}(b) shows the time-resolved emission of an OPO  excited in same trap by tuning the pump laser energy to the one from the second confined level at low excitation conditions (right panel).
The signal blue-shift in the OPO configuration is negligible compared to the one under  the non-resonant excitation [cf.~Fig.~\ref{FigTOPO}(a)] due to the absence of an excitonic reservoir.  
The emission from  the signal  and  idler OPO states only appears during the restricted range of SAW phases  for which these states are equidistant to the pump. The time dependence  confirms the dynamic nature of the acoustic tuning. The turn-on and turn-off times of the OPO (taken as the time delay for the intensity of the signal state to change by an order of magnitude) is far below the temporal resolution of the present measurements of approx. 100~ps. 

\begin{figure}[t!bhp]
    \centering
    \includegraphics[width=1\columnwidth, angle=0, clip]{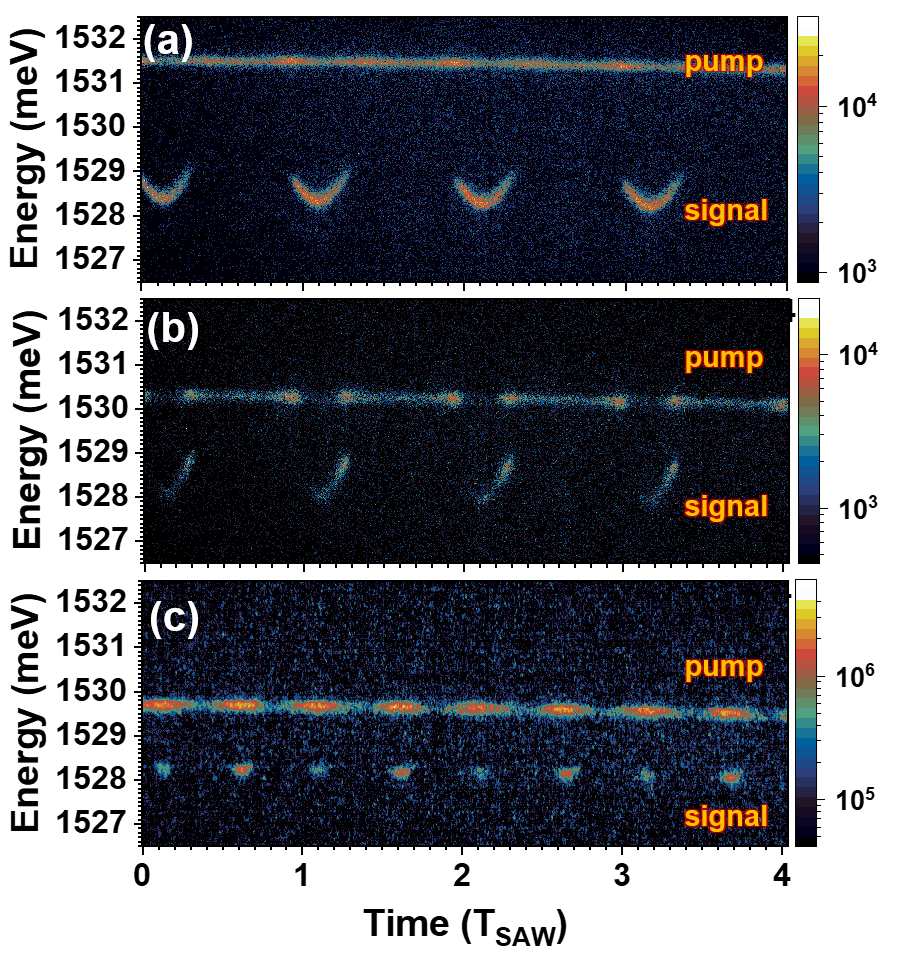}
    \caption{{\bf OPO at different pump conditions}. 
Time-resolved PL maps of an acoustically triggered OPO in a nominally  $4\times4~\mu$m$^2$ trap excited by pump energies (a) 1531.5~meV, (b) 1530.3~meV, and (c) 1529.7~meV. Note that  in (c) the OPO triggers twice in a SAW period  ($\TSAW=2.61$ ns).
    \label{FigPumpEnergy}}
\end{figure}

The OPO triggering dynamics depends on the SAW amplitude and phase, as well as on  the pump energy. Figure~\ref{FigPumpEnergy} illustrates the different dynamic regimes that can be induced by varying the pump energy.  At low pump energies, the OPO is normally excited only over a small range of SAW phases. Depending on the SAW amplitude and pump energy, the OPO can be triggered once in a SAW cycle, as in Fig.~\ref{FigPumpEnergy}(a \& b),  or twice in  in a SAW cycle, as shown in Fig.~\ref{FigPumpEnergy}(c). For high pump energies, the OPO can remain triggered over a range of SAW phases -- the latter enables us to follow the dynamic energy modulation of the signal state induced by the SAW field, as illustrated in Figs.~\ref{FigPumpEnergy}(a) and \ref{FigPumpEnergy}(b).  Note that in all cases, the signal energy for the OPO triggering is always  slightly lower than the one for the OPO turnoff, thus showing a hysteretic dependence on the SAW phase. This behavior is attributed to the dynamic energy shifts of the OPO states arising from polariton-polariton interactions. Once the OPO is triggered, the SAW-induced energy shifts of the OPO levels can be counteracted by changes in the polariton density induced by the stimulated scattering to the signal and idler states. In this way, it becomes possible to fulfill the OPO energy conservation requirement for a range of SAW phases and energies of the pump state.

\begin{figure*}[t!bhp]
    \includegraphics[width=0.9\textwidth, keepaspectratio=true]{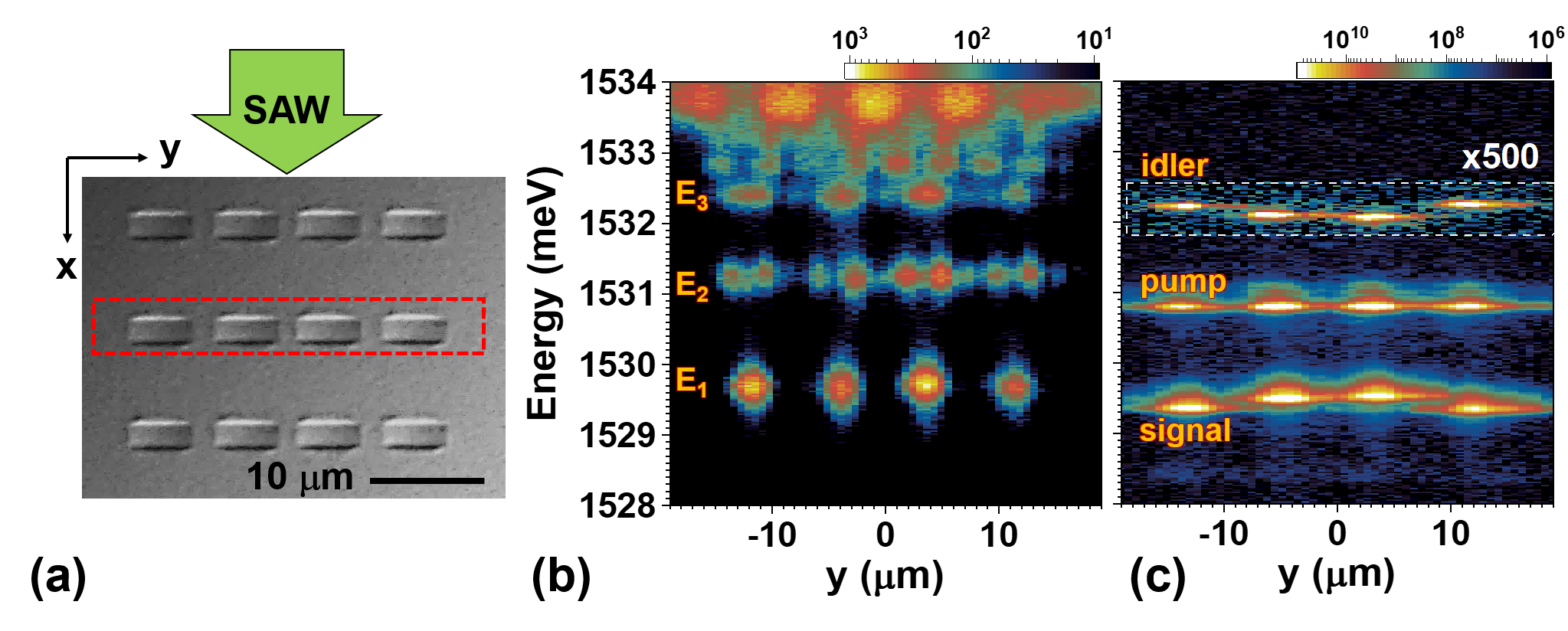}
    \caption{{\bf Tuned arrays of acoustically tuned OPOs}. 
 (a) Optical micrograph of an array of nominally $3\times3~\mu$m$^2$ intra-cavity traps with a pitch of $9~\mu$m. The dashed rectangle designates the region probed by PL.   (b)-(c)  Spectral PL map of the array recorded  (a)  under weak non-resonant excitation in the absence of SAW, and (b) resonant excitation in the presence of SAW. 
}
    \label{FigArray}
\end{figure*}

\subsection{OPO arrays}
\label{OPO_arrays}

We now demonstrate that the acoustic triggering is very robust against fluctuations in trap size and energies, thus making it possible to synchronously trigger OPOs in an array of traps. The studies were carried out in a square array of $3\times3~\mu$m$^2$ traps with a pitch of $9~\mu$m  schematically depicted in Fig.~\ref{FigArray}(a). Figure~\ref{FigArray}(b) displays a PL map, recorded under weak non-resonant excitation and absence of SAW by collecting the PL within the area indicated by the dashed square in the Fig.~\ref{FigArray}(a). The large spatial separation between the traps prevents tunnel coupling between them. The map thus reveals a series of confined states with almost identical energy spectrum for all traps. OPO experiment conditions are identical to the single $4\times4~\mu$m$^2$ trap discussed above. 
Figure~\ref{FigArray}(c) shows the corresponding PL map obtained by pumping the array slightly below the $\Psi_\mathrm{12}$ level in the presence of the acoustic field. OPO is activated in all intra-cavity traps with correlated triggering times determined by the SAW phase at the trap location. Similar to the single trap [cf. Fig.~\ref{FigOPO}(c)], no OPO excitation is observed in the absence of a SAW. The variations of the  signal and idler energies on trap position arises from the Gaussian shape of the exciting laser beam, which populates the traps with different polariton densities. These fluctuations, however, do not prevent OPO triggering in all lattice sites under the acoustic modulation.

\section{\label{Discussions}Discussions}


\subsection{Acoustic modulation of confined levels}
\label{Acoustic modulation of confined levels}

The state dependent acoustic energy tuning mainly relies  on the SAW strain field. The latter modulates the excitonic component of polariton via the deformation potential mechanism as well as the photonic component due to modulation of the thicknesses and refractive indices of the MC layers \cite{PVS107,PVS156}. These two modulation mechanisms add  in phase for the present sample and lead to the effective standing  polariton potential given by Eq.~(\ref{EqVSAW}).

The symmetry and time evolution of the polariton levels \dAK{\cPVS{first appearance!!!}{in Fig.~\ref{FigOPO}(e)}} under a standing acoustic fields
can be understood by using perturbation theory to calculate the impact of the dynamical potential  $V_\mathrm{SAW}$ [cf.~Eq.~(\ref{EqWF})] on the   confined levels given by Eq.~(\ref{EqE}). 
The acoustically induced energy shifts of the $\Psi_\mathrm{m_xm_y}$ can be expressed as (see Sec.~\ref{SM:Acoustic_Modulation} for details):

\begin{eqnarray}
\Delta E_\mathrm{m_x m_y} &=& \langle\Psi_\mathrm{m_xm_y}|\phi_\mathrm{SAW}|\Psi_\mathrm{m_xm_y} \rangle\\ \nonumber
&\approx& V_\mathrm{SAW,0}   
\left( 
	\frac{\pi^2}{6} - \frac{1}{m_x^2}
\right) 
\left(\frac{l_x}{\lSAW}\right)^2 \cos{(\pSAW)}  E_\mathrm{11} 
\label{EqSIDESAW}
\end{eqnarray}

The acoustic modulation thus introduces a  spatial distortion of the confinement potential, which is dictated by the instantaneous amplitude $V_\mathrm{SAW,0}\cos{(\pSAW)}$ of the standing field as well as by the ratio between the trap dimensions and the acoustic wavelength. The corresponding energy shifts $\Delta E_\mathrm{m_x m_y}$ are independent of the  mode index $m_y$ but reduce with increasing $m_x$. This behavior  arises from the fact that the $m_y$ lobes of the wave function are centered on the SAW anti-nodes, thus experiencing the full strain-induced  energy modulation. The $m_x$ lobes, in contrast, are oriented along the SAW propagation direction and, thus, probe different phases of the standing SAW field. 
These state-dependent shifts enable the dynamic energy tuning  for OPO triggering over a wide range of trap geometries (see Sec.~\ref{SM:Acoustic_Modulation}). In particular, the OPO states $O_2=\{\Psi_{11}, \Psi_{21}, \Psi_{31}\}$ in a perfect square trap become equidistant in energy by selecting the SAW amplitude and phase to satisfy:

\begin{equation}
V_\mathrm{SAW,0}\cos{(\pSAW)} = \frac{2}{3} \left(\frac{\lSAW}{\ell}\right)^2
\label{EqT2}
\end{equation}
In order to determine the nature of the OPO states, we first examine the impact of the SAW on the energy of the  states  given by Eq.~(\ref{EqWF}).  Figure~\ref{FigOPO}(e) displays the energy evolution of the confined polariton levels with the SAW phase, $\pSAW$. The calculations were carried out using a numerical approach that takes into account the measured spatial profile of the traps (see Sec.~\ref{SM:Intra-cavity_polariton_traps} for details), but neglects polariton-interactions. A more realistic model taking into account interactions will be presented in Sec.~\ref{Numerical simulations of the OPO dynamics}. The vertical dash-dotted black lines mark the nodes of the SAW strain field, where the polariton states are identical to the ones of an unperturbed trap. The states within set $O_1=\{\Psi_{11}, \Psi_{12} \enskip\text{or}\enskip \Psi_{21}, \Psi_{22}\}$ are approximately equidistant at this phase, but, for symmetry reasons presented in Sec.~\ref{Symmetry of the OPO states}, do not interact\dAK{to form an OPO even under a SAW}. The vertical dashed blue lines indicate the phases for which the OPO energy matching requirement becomes satisfied for the set $O_2=\{\Psi_{11}, \Psi_{21} , \Psi_{31}\//\Psi_{13}\}$. We attribute the PL features in Fig.~\ref{FigOPO}(d)  to an OPO involving these states. This assignment is supported by a comparison of the PL maps with their calculated wave squared function  $|\Psi_\mathrm{m_x m_y}|^2$ 
at the matching SAW phase displayed in Figs.~\ref{FigOPO}(f)-(h). The pump state has thus  a predominantly $\Psi_\mathrm{21}$ character with two lobes along $x$. The idler state results from the SAW-induced red-shift of the $\Psi_\mathrm{13}$ unperturbed state, which  mixes with the $\Psi_{31}$ state. The idler state, thus acquires  the symmetry shown in Fig.~\ref{FigOPO}(f). 

When the pump energy is in resonance or blue-shifted with respect to the $\Psi_{21}$ state, the OPO assumes the $O_3$ configuration (cf. Fig.~\ref{FigOPOb}) with an $\Psi_{31}$ idler state.

\subsection{Symmetry of the OPO states}
\label{Symmetry of the OPO states}

The previous sections have shown that acoustic tuning enables the excitation of OPO in intra-cavity traps over a wide range optical and acoustical excitation conditions. One interesting question is why it is not possible to trigger an OPO in the configuration $O_1=\{\Psi_{11}, \Psi_{21} (\text{or } \Psi_{12}), \Psi_{22}\}$, which has equally separated states for a perfect square potential. In fact, we show in the SM (Sec.~\ref{Acoustically_induced_energy_tunning}) that deviations from a square shape can also be corrected  by the acoustic field. The required field amplitudes are in this case much smaller than those  given by Eq.~(\ref{EqT2}) for the energy matching of the $O_2$ states. 

The inability to excite an $O_1$  OPO  arises from  symmetry requirements of  the non-linear process responsible for OPO triggering, a critical process to initiate parametric oscillations. OPO triggering initiates when fluctuations in population leads to the occupation of the pump state by two polaritons. The latter creates a non-adiabatic and non-linear potential that  couples the initial two-polariton state $\{\Psi_p\Psi_p\}$ to  the final state   $\{\Psi_\mathrm{s}\Psi_\mathrm{i}\}$ consisting of particles in a superposition of signal and idler states. In the contact approximation for polariton-polariton interactions,\cite{Ciuti_PRB63_41303_01,Carusotto_PRB72_125335_05} the perturbed two-polariton state   represented by $\{\Psi_p\Psi_p\}'$ can be expressed as:

\begin{equation}
\label{EqNL}
\{\Psi_p\Psi_p\}'  \approx \{\Psi_p\Psi_p\} + 
   \frac{ \langle \{\Psi_\mathrm{s}\Psi_\mathrm{i}\} | \delta V_p | \{\Psi_\mathrm{p}\Psi_\mathrm{p}\}\rangle}{2(E_\mathrm{p}-E_\mathrm{s})}  \{\Psi_s\Psi_i\} +\dots
\end{equation}

\noindent The coupling Hamiltonian $\delta V_p$ can be expressed as

\begin{equation}
\label{deltaVP}
\delta V_p=\int{ g \left[ \Psi_\mathrm{i} \Psi_\mathrm{s} \Psi_\mathrm{p}^2\right]} dx dy,
\end{equation}

\noindent where $g$ denotes the effective polariton-polariton coupling strength. The non-linear coupling $\delta V_p$ thus enables the scattering of pump polaritons to the idler and signal states required to trigger the stimulated scattering leading to parametric oscillations and amplification.

Equation~(\ref{deltaVP}) has important consequences for potentials with mirror symmetry, such as the ones created by the intra-cavity traps studied here. The confined states $\Psi_{ij}$ in these potentials have a well-defined parity. Since  $|\Psi_\mathrm{p}|^ 2$  in Eq.~\ref{deltaVP} is always an even function,  a non-vanishing coupling  $\delta V_p$ requires idler and signal states with same parity,  i.e., states with indices $m_i$ ($i=x,y$) differing by an even number. This condition is satisfied for the sets $O_2$ and $O_3$ but not for $O_1$, in full agreement with the experimental results.

The previously mentioned symmetry requirements  remain valid under a standing SAW field, as long as the traps are centered at a field anti-node, since in this case  the acoustic perturbation  does not affect the mirror symmetry of the trap potential. If, however, the SAW field has a traveling component (or if the trap is displaced from the anti-nodes of a standing field), it will mix states with different parities  and  enable other  OPO configurations.

\begin{figure*}[tbhp]
    \includegraphics[width=1.0\textwidth, angle=0, clip]{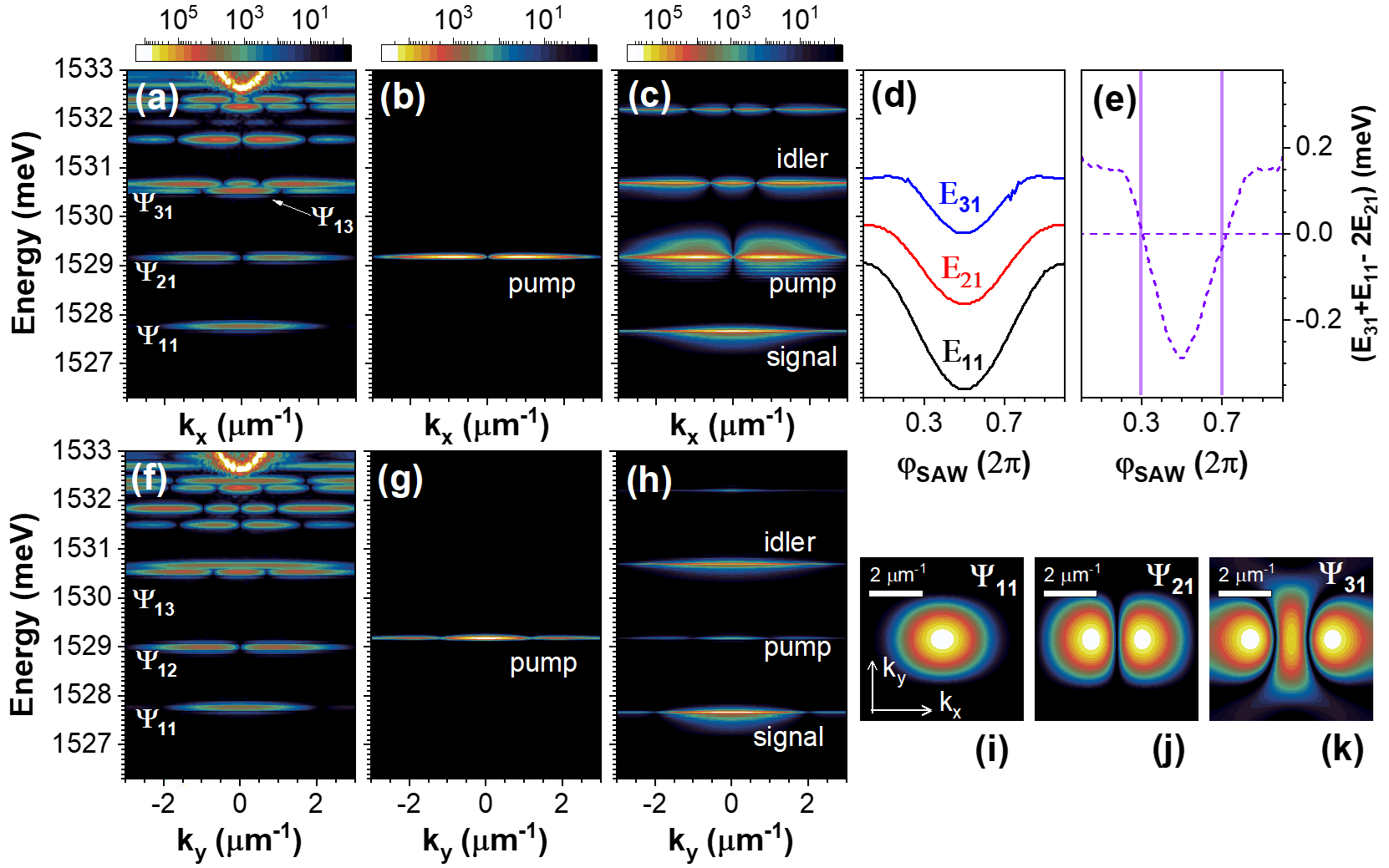}
    \caption{
   {\bf  Theoretical analysis of  acoustically tuned OPOs}. 
Numerical simulations carried out by solving the driven-dissipative Gross-Pitaevskii equation (GPE) for polaritons with an $4\times4~\mu$m$^2$ intra-cavity trap. 
    (a) and (f) Momentum-space spectra of the trap under non-resonant excitation and in the absence of SAW along x and y directions, respectively. (b) and (g) Momentum-space spectra of the trap under resonant excitation into  $\Psi_\mathrm{21}$ state in the absence of SAW along x and y directions, respectively. (c) and (h) Similar to (b) and (g) but for the SAW phase of 0.3 in units of $2\pi$. (d) Evolution of the energy of the OPO levels over one SAW cycle. (e) Energy difference ($\Delta E=E_{31}+E_{11}-2E_{21}$) between the  $\Psi_\mathrm{ij}$  levels over once SAW period. The vertical bars designate two SAW phases for which the spectrum of the trap is equidistant ($\Delta E = 0$). (i-k) Spatial profiles of the squared wavefunctions of the signal, pump and idler states at SAW phase of $0.3(2\pi)$.}
    \label{SimUCL}
\end{figure*}

\subsection{Driven-dissipative simulations of the OPO dynamics}
\label{Numerical simulations of the OPO dynamics}

We present in this section a theoretical analysis based on the numerical solution of the driven-dissipative Gross-Pitaevskii equation (GPE) for the lower-polariton field $(\Psi)$ in an intra-cavity trap subjected to an optical field ($ F_\mathrm{p}(\textbf{r},t)$ with $\textbf{r}=(x,y)$) as well as to the  acoustic modulation potential given by Eq.~(\ref{EqVSAW}). In contrast to the single-particle calculations presented above, the GPE solutions implicitly account for the  polariton non-linearity leading to OPO formation.  In atomic units,  the GPE can be expressed as:

\begin{align}
\label{EqGPE}
&i\partial_{t}\Psi(\textbf{r},t) = F_\mathrm{p}(\textbf{r},t)  + \left(\right. \omega_\mathrm{LP}(-i\nabla)-i\kappa_\mathrm{LP} \\ \nonumber
&+g_\mathrm{LP}|\Psi(\textbf{r},t)|^2+V_\mathrm{trap}(\textbf{r})+V_\mathrm{SAW}(\textbf{r},t)\left.\right)\Psi(\textbf{r},t) .
\end{align}

\noindent Here, $\omega_\mathrm{LP}$, $\kappa_\mathrm{LP}$ and $g_\mathrm{LP}$ are lower-polariton dispersion, decay rate and polariton-polariton interaction, respectively. The dimensions of the  intra-cavity trap were determined from AFM height maps, see \dAK{ Eq.~(\ref{SM:Eqdz}) and }Table~\ref{tab:SMT1}. The parameters for the trap potential and the SAW modulation amplitude were the same as in the experiments. More details of the calculations procedure are summarized in Sec.~\ref{Numerical Simulation of the acoustically driven OPO}.

Figures~\ref{SimUCL}(a) and \ref{SimUCL}(f) display the wave-function projections of the polariton states (in  momentum space) calculated  for low power, non-resonant optical excitation of the trap.\dAK{The plots reproduce very well the corresponding PL maps in Fig.~\ref{FigOPOb}(b) (except for a constant shift of the energy axis).} The central panels [Figs.~\ref{SimUCL}(b) and \ref{SimUCL}(g)] show the corresponding maps under polariton injection into the $\Psi_\mathrm{21}$ state, which blue-shifts due to the polariton-polariton interactions. In agreement with the experimental results of in the absence of a SAW, no OPO is observed under these conditions. In contrast, once the SAW potential is added, the signal and idler states appear in the simulated spectra [cf. Figs.~\ref{SimUCL}(c) and \ref{SimUCL}(h)]. We emphasize here the excellent agreement between the calculated wave function projection of Fig.~\ref{SimUCL}(h) and the  experimental momentum-resolved PL map of Fig.~\ref{FigOPOb}(b).

The wave functions of the OPO states are illustrated in Figs.~\ref{SimUCL}(i)-(k) correspond to the ones expected for an OPO involving the state set  $O_3=\{\Psi_{11}, \Psi_{21} , \Psi_{31}\}$. The dependence of their energy  on the SAW phase, which is summarized in Figs.~\ref{SimUCL}(d)  and \ref{SimUCL}(e), shows that the OPO energy matching condition becomes satisfied twice in a SAW cycle for $\pSAW=0.3$ and 0.7~rad.
\dAK{Here, the energy matching arises from the fact that the strain-induced energy modulation on the $\Psi_\mathrm{m_x1}$ states decreases with increasing index $m_x$ (cf.~Eq.~\ref{EqSIDESAW}). As a result, the matching is achieved when the SAW strain blueshifts the confined levels (in contrast to the $O_2$ OPO of Fig.~\ref{FigOPO}, which triggers during a compressive cycle of the SAW field).}

\subsection{Idler-signal intensity ratios}
\label{Idler-signal intensity ratios}

The OPO process yields pairs of signal and idler polaritons: the same applies for the  simulation of the Fig.~\ref{SimUCL}(c,h) that predicts almost identical amplitudes for the signal and idler states. The PL yield from these states depends on how the polaritons decay to photons and may differ considerably due to differences in the  scattering rates, Hopfield coefficients, emission pattern, and photon re-absorption. As a result, the emission from idler states is normally much weaker than the signal one. The latter is a main drawback for applications as sources of correlated photons, which  ideally require comparable intensity ratios.  

The ratio $r_\mathrm{OPO}$ between the integrated emission intensity of the signal and idler in the present studies covers a wide range extending from 20 to 500. For comparison, the OPOs based on confined polariton states reported in Ref.~\cite{Ferrier_APL97_31105_10} have $r_\mathrm{OPO}$ ratios ranging from approx. 5 to 100, depending on the excitation intensity, which compare with the range from 5 to 10 predicted by theoretical studies presented in the same manuscript. The same ratio increases to  $10^3$ to $10^4$  in polariton OPOs based on tripple microcavities \cite{Diederichs_N440_904_06}. 

The high $r_\mathrm{OPO}$ inferred from Figs.~\ref{FigOPO}-\ref{FigArray}  is  partially due to an inefficient collection of the idler emission. In particular, the emission in the PL maps will appear very weak if the idler state has a spatially-extended wave function or,  as discussed in Sec.~\ref{OPO in intra-cavity traps},  an emission node along the collection axis. We estimate that the limited collection of the idler emission, which can be eliminated by a full measurement of the wave functions, induces an increases of the measured  $r_\mathrm{OPO}$ by a factor between 3 and 5.

Another mechanism leading to a large  $r_\mathrm{OPO}$ arises from the higher photonic  content of the signal states in comparison to the idler states in the present sample. The  energy difference between signal and idler states is less than half of the Rabi splitting, so that the large $r_\mathrm{OPO}$ ratios can not be solely attributed to differences in the Hopfield coefficients. 
Finally, the large $r_\mathrm{OPO}$ may also arise from decay paths from the pump to the low lying signal states  (e.g., by a thermal process).  Figure~\ref{FigOPO}(d) shows, however, that the signal state only emits under OPO excitation, thus proving the absence of a parallel excitation path. We suggest that the weak photon yield of the idler results from the strong dephasing arising from the coupling with closely lying energy levels. The acoustic field may play a role in this process: in particular, the SAWs employed here also carry piezoelectric fields, which interact strongly and can efficiently mix electronic states. Future acoustic modulation studies using non-piezoelectric SAWs \cite{PVS177} will help to clarify this issue.

\section{\label{Conclusions}Summary and Outlook}
 
We have  demonstrated  an efficient and versatile approach for the dynamic control of the scattering pathways of confined polariton condensates based on the modulation by spatially and time-varying potentials produced by SAWs. A unique feature arising from the spatial dependence of the SAW field  is the ability to dynamically control the energy of individual polariton states in a confined potential. Here, the SAW is applied to  tune the energy states of confined exciton-polariton condensates to enable OPOs. We demonstrated that the acoustic OPO requires not only the matching of the energy energy-level separation, but also signal, pump, and idler states with the appropriate symmetry. The experimental studies have been complemented by a theoretical framework, which accounts for the required symmetry of the confined states and also provides a quantitative determination of the energy tuning parameters.
Finally, we have presented experimental results confirming the dynamic character and the  robust nature of the acoustic tuning, which enables OPOs under  a wide range of excitation conditions.  

A natural future step will be the exploitation of  acoustically tuned  OPOs for the generation of entangled photons from a single trap and arrays. We anticipate that one of the challenges will be to control the mismatch in emission intensity between  the signal and idler states. \dAK{discussed in Sec.~\ref{Idler-signal intensity ratios}.} The theoretical framework together with the ability to develop polariton confinement potentials with the appropriate symmetry  are an excellent starting point to reach this goal. 

The ability to synchronously tune several OPOs is one further advantage of the dynamic acoustic tuning. This functionality is demonstrated by the excitation of an array of confined OPOs using a single acoustic beam. The modulation of the individual polariton traps at the array nodes appropriately tunes the confined levels and counteracts unavoidable energy fluctuations. Furthermore, the OPO emission from the array sites is correlated by the SAW phase. The time jitter of the emission depends on the fluctuations in the trap properties and can be minimized by increasing the SAW amplitude. The photon pairs are emitted not only at well-defined locations within the array but also at well-determined times, a feature which can enhance the  fidelity of such a source of correlated photons.

As a final remark, we point out that strain fields  interact with a wide variety of excitations in solid state systems. The  dynamical acoustic tuning reported can  thus be  applied to a wide variety of systems, thus providing the robustness in operation required for  the realization of scalable on-chip systems.

{\noindent {\it Acknowledgements: }} 
We thank M. Ramsteiner and S. Krishnamurthy  for discussions and for a critical review of the manuscript. We also acknowledge the technical support from R. Baumann, S. Rauwerdink, and A. Tahraoui in the sample fabrication process.  We acknowledge financial support from the German DFG (grant 359162958) and from the QuantERA grant Interpol (EU-BMBF (Germany) grant nr. 13N14783).

\IfFileExists{literature.bib}
	{ \def\litdir{.} }
{ \IfFileExists{x:/sawoptik_databases/jabref/literature.bib}  
	{  \def\litdir{x:/sawoptik_databases/jabref} }
      {   \def\litdir{c:/myfiles/jabref}     }
}


\begin{thebibliography}{24}%
\makeatletter
\providecommand \@ifxundefined [1]{%
 \@ifx{#1\undefined}
}%
\providecommand \@ifnum [1]{%
 \ifnum #1\expandafter \@firstoftwo
 \else \expandafter \@secondoftwo
 \fi
}%
\providecommand \@ifx [1]{%
 \ifx #1\expandafter \@firstoftwo
 \else \expandafter \@secondoftwo
 \fi
}%
\providecommand \natexlab [1]{#1}%
\providecommand \enquote  [1]{``#1''}%
\providecommand \bibnamefont  [1]{#1}%
\providecommand \bibfnamefont [1]{#1}%
\providecommand \citenamefont [1]{#1}%
\providecommand \href@noop [0]{\@secondoftwo}%
\providecommand \href [0]{\begingroup \@sanitize@url \@href}%
\providecommand \@href[1]{\@@startlink{#1}\@@href}%
\providecommand \@@href[1]{\endgroup#1\@@endlink}%
\providecommand \@sanitize@url [0]{\catcode `\\12\catcode `\$12\catcode
  `\&12\catcode `\#12\catcode `\^12\catcode `\_12\catcode `\%12\relax}%
\providecommand \@@startlink[1]{}%
\providecommand \@@endlink[0]{}%
\providecommand \url  [0]{\begingroup\@sanitize@url \@url }%
\providecommand \@url [1]{\endgroup\@href {#1}{\urlprefix }}%
\providecommand \urlprefix  [0]{URL }%
\providecommand \Eprint [0]{\href }%
\providecommand \doibase [0]{https://doi.org/}%
\providecommand \selectlanguage [0]{\@gobble}%
\providecommand \bibinfo  [0]{\@secondoftwo}%
\providecommand \bibfield  [0]{\@secondoftwo}%
\providecommand \translation [1]{[#1]}%
\providecommand \BibitemOpen [0]{}%
\providecommand \bibitemStop [0]{}%
\providecommand \bibitemNoStop [0]{.\EOS\space}%
\providecommand \EOS [0]{\spacefactor3000\relax}%
\providecommand \BibitemShut  [1]{\csname bibitem#1\endcsname}%
\let\auto@bib@innerbib\@empty
\bibitem [{\citenamefont {Weisbuch}\ \emph {et~al.}(1992)\citenamefont
  {Weisbuch}, \citenamefont {Nishioka}, \citenamefont {Ishikawa},\ and\
  \citenamefont {Arakawa}}]{Weisbuch92a}%
  \BibitemOpen
  \bibfield  {author} {\bibinfo {author} {\bibfnamefont {C.}~\bibnamefont
  {Weisbuch}}, \bibinfo {author} {\bibfnamefont {M.}~\bibnamefont {Nishioka}},
  \bibinfo {author} {\bibfnamefont {A.}~\bibnamefont {Ishikawa}},\ and\
  \bibinfo {author} {\bibfnamefont {Y.}~\bibnamefont {Arakawa}},\ }\bibfield
  {title} {\bibinfo {title} {Observation of the coupled exciton-photon mode
  splitting in a semiconductor quantum microcavity},\ }\href@noop {} {\bibfield
   {journal} {\bibinfo  {journal} {Phys. Rev. Lett.}\ }\textbf {\bibinfo
  {volume} {69}},\ \bibinfo {pages} {3314} (\bibinfo {year}
  {1992})}\BibitemShut {NoStop}%
\bibitem [{\citenamefont {Kasprzak}\ \emph {et~al.}(2006)\citenamefont
  {Kasprzak}, \citenamefont {Richard}, \citenamefont {Kundermann},
  \citenamefont {Baas}, \citenamefont {Jeambrun}, \citenamefont {Keeling},
  \citenamefont {Marchetti}, \citenamefont {Szyma\'nska}, \citenamefont
  {Andr{\'e}}, \citenamefont {Staehli}, \citenamefont {Savona}, \citenamefont
  {Littlewood}, \citenamefont {Deveaud},\ and\ \citenamefont
  {Dang}}]{Kasprzak_N443_409_06}%
  \BibitemOpen
  \bibfield  {author} {\bibinfo {author} {\bibfnamefont {J.}~\bibnamefont
  {Kasprzak}}, \bibinfo {author} {\bibfnamefont {M.}~\bibnamefont {Richard}},
  \bibinfo {author} {\bibfnamefont {S.}~\bibnamefont {Kundermann}}, \bibinfo
  {author} {\bibfnamefont {A.}~\bibnamefont {Baas}}, \bibinfo {author}
  {\bibfnamefont {P.}~\bibnamefont {Jeambrun}}, \bibinfo {author}
  {\bibfnamefont {J.~M.~J.}\ \bibnamefont {Keeling}}, \bibinfo {author}
  {\bibfnamefont {F.~M.}\ \bibnamefont {Marchetti}}, \bibinfo {author}
  {\bibfnamefont {M.~H.}\ \bibnamefont {Szyma\'nska}}, \bibinfo {author}
  {\bibfnamefont {R.}~\bibnamefont {Andr{\'e}}}, \bibinfo {author}
  {\bibfnamefont {J.~L.}\ \bibnamefont {Staehli}}, \bibinfo {author}
  {\bibfnamefont {V.}~\bibnamefont {Savona}}, \bibinfo {author} {\bibfnamefont
  {P.~B.}\ \bibnamefont {Littlewood}}, \bibinfo {author} {\bibfnamefont
  {B.}~\bibnamefont {Deveaud}},\ and\ \bibinfo {author} {\bibfnamefont {L.~S.}\
  \bibnamefont {Dang}},\ }\bibfield  {title} {\bibinfo {title} {{Bose-Einstein}
  condensation of exciton polaritons},\ }\href
  {https://doi.org/10.1038/nature05131} {\bibfield  {journal} {\bibinfo
  {journal} {Nature}\ }\textbf {\bibinfo {volume} {443}},\ \bibinfo {pages}
  {409} (\bibinfo {year} {2006})}\BibitemShut {NoStop}%
\bibitem [{\citenamefont {Carusotto}\ and\ \citenamefont
  {Ciuti}(2013)}]{Carusotto_RMP85_299_13}%
  \BibitemOpen
  \bibfield  {author} {\bibinfo {author} {\bibfnamefont {I.}~\bibnamefont
  {Carusotto}}\ and\ \bibinfo {author} {\bibfnamefont {C.}~\bibnamefont
  {Ciuti}},\ }\bibfield  {title} {\bibinfo {title} {Quantum fluids of light},\
  }\href@noop {} {\bibfield  {journal} {\bibinfo  {journal} {Rev. Mod. Phys.}\
  }\textbf {\bibinfo {volume} {85}},\ \bibinfo {pages} {299} (\bibinfo {year}
  {2013})}\BibitemShut {NoStop}%
\bibitem [{\citenamefont {Sanvitto}\ and\ \citenamefont
  {Kena-Cohen}(2016)}]{Sanvitto_NM15_1061_16}%
  \BibitemOpen
  \bibfield  {author} {\bibinfo {author} {\bibfnamefont {D.}~\bibnamefont
  {Sanvitto}}\ and\ \bibinfo {author} {\bibfnamefont {S.}~\bibnamefont
  {Kena-Cohen}},\ }\bibfield  {title} {\bibinfo {title} {The road towards
  polaritonic devices},\ }\href {http://dx.doi.org/10.1038/nmat4668} {\bibfield
   {journal} {\bibinfo  {journal} {Nat Mater}\ }\textbf {\bibinfo {volume}
  {15}},\ \bibinfo {pages} {1061} (\bibinfo {year} {2016})}\BibitemShut
  {NoStop}%
\bibitem [{\citenamefont {Savvidis}\ \emph {et~al.}(2000)\citenamefont
  {Savvidis}, \citenamefont {Baumberg}, \citenamefont {Stevenson},
  \citenamefont {Skolnick}, \citenamefont {Whittaker},\ and\ \citenamefont
  {Roberts}}]{Savvidis_PRL84_1547_00}%
  \BibitemOpen
  \bibfield  {author} {\bibinfo {author} {\bibfnamefont {P.~G.}\ \bibnamefont
  {Savvidis}}, \bibinfo {author} {\bibfnamefont {J.~J.}\ \bibnamefont
  {Baumberg}}, \bibinfo {author} {\bibfnamefont {R.~M.}\ \bibnamefont
  {Stevenson}}, \bibinfo {author} {\bibfnamefont {M.~S.}\ \bibnamefont
  {Skolnick}}, \bibinfo {author} {\bibfnamefont {D.~M.}\ \bibnamefont
  {Whittaker}},\ and\ \bibinfo {author} {\bibfnamefont {J.~S.}\ \bibnamefont
  {Roberts}},\ }\bibfield  {title} {\bibinfo {title} {Angle-resonant stimulated
  polariton amplifier},\ }\href {https://doi.org/10.1103/PhysRevLett.84.1547}
  {\bibfield  {journal} {\bibinfo  {journal} {Phys. Rev. Lett.}\ }\textbf
  {\bibinfo {volume} {84}},\ \bibinfo {pages} {1547} (\bibinfo {year}
  {2000})}\BibitemShut {NoStop}%
\bibitem [{\citenamefont {Saba}\ \emph {et~al.}(2001)\citenamefont {Saba},
  \citenamefont {Ciuti}, \citenamefont {Bloch}, \citenamefont {Thierry-Mieg},
  \citenamefont {Andr\'{e}}, \citenamefont {Dang}, \citenamefont {Kundermann},
  \citenamefont {Mura}, \citenamefont {Bongiovanni}, \citenamefont {Staehli},\
  and\ \citenamefont {Deveaud}}]{Saba01a}%
  \BibitemOpen
  \bibfield  {author} {\bibinfo {author} {\bibfnamefont {M.}~\bibnamefont
  {Saba}}, \bibinfo {author} {\bibfnamefont {C.}~\bibnamefont {Ciuti}},
  \bibinfo {author} {\bibfnamefont {J.}~\bibnamefont {Bloch}}, \bibinfo
  {author} {\bibfnamefont {V.}~\bibnamefont {Thierry-Mieg}}, \bibinfo {author}
  {\bibfnamefont {R.}~\bibnamefont {Andr\'{e}}}, \bibinfo {author}
  {\bibfnamefont {S.~S.}\ \bibnamefont {Dang}}, \bibinfo {author}
  {\bibfnamefont {S.}~\bibnamefont {Kundermann}}, \bibinfo {author}
  {\bibfnamefont {A.}~\bibnamefont {Mura}}, \bibinfo {author} {\bibfnamefont
  {G.}~\bibnamefont {Bongiovanni}}, \bibinfo {author} {\bibfnamefont {J.~L.}\
  \bibnamefont {Staehli}},\ and\ \bibinfo {author} {\bibfnamefont
  {B.}~\bibnamefont {Deveaud}},\ }\bibfield  {title} {\bibinfo {title}
  {High-temperature ultrafast polariton parametric amplification in
  semiconductor microcavities},\ }\href@noop {} {\bibfield  {journal} {\bibinfo
   {journal} {Nature}\ ,\ \bibinfo {pages} {731}} (\bibinfo {year}
  {2001})}\BibitemShut {NoStop}%
\bibitem [{\citenamefont {Baumberg}\ \emph {et~al.}(2000)\citenamefont
  {Baumberg}, \citenamefont {Savvidis}, \citenamefont {Stevenson},
  \citenamefont {Tartakovskii}, \citenamefont {Skolnick}, \citenamefont
  {Whittaker},\ and\ \citenamefont {Roberts}}]{Baumberg00a}%
  \BibitemOpen
  \bibfield  {author} {\bibinfo {author} {\bibfnamefont {J.~J.}\ \bibnamefont
  {Baumberg}}, \bibinfo {author} {\bibfnamefont {P.~G.}\ \bibnamefont
  {Savvidis}}, \bibinfo {author} {\bibfnamefont {R.~M.}\ \bibnamefont
  {Stevenson}}, \bibinfo {author} {\bibfnamefont {A.~I.}\ \bibnamefont
  {Tartakovskii}}, \bibinfo {author} {\bibfnamefont {M.~S.}\ \bibnamefont
  {Skolnick}}, \bibinfo {author} {\bibfnamefont {D.~M.}\ \bibnamefont
  {Whittaker}},\ and\ \bibinfo {author} {\bibfnamefont {J.~S.}\ \bibnamefont
  {Roberts}},\ }\bibfield  {title} {\bibinfo {title} {Parametric oscillation in
  a vertical microcavity: A polariton condensate or micro-optical parametric
  oscillation},\ }\href@noop {} {\bibfield  {journal} {\bibinfo  {journal}
  {Phys. Rev. B}\ }\textbf {\bibinfo {volume} {62}},\ \bibinfo {pages} {R16247}
  (\bibinfo {year} {2000})}\BibitemShut {NoStop}%
\bibitem [{\citenamefont {Langbein}(2004)}]{Langbein_PRB70_205301_04}%
  \BibitemOpen
  \bibfield  {author} {\bibinfo {author} {\bibfnamefont {W.}~\bibnamefont
  {Langbein}},\ }\bibfield  {title} {\bibinfo {title} {Spontaneous parametric
  scattering of microcavity polaritons in momentum space},\ }\href
  {https://doi.org/10.1103/PhysRevB.70.205301} {\bibfield  {journal} {\bibinfo
  {journal} {Phys Rev B}\ }\textbf {\bibinfo {volume} {70}},\ \bibinfo {pages}
  {205301} (\bibinfo {year} {2004})}\BibitemShut {NoStop}%
\bibitem [{\citenamefont {Ciuti}(2004)}]{Ciuti_PRB69_245304_04}%
  \BibitemOpen
  \bibfield  {author} {\bibinfo {author} {\bibfnamefont {C.}~\bibnamefont
  {Ciuti}},\ }\bibfield  {title} {\bibinfo {title} {Branch-entangled polariton
  pairs in planar microcavities and photonic wires},\ }\href@noop {} {\bibfield
   {journal} {\bibinfo  {journal} {Phys. Rev. B}\ }\textbf {\bibinfo {volume}
  {69}},\ \bibinfo {pages} {245304} (\bibinfo {year} {2004})}\BibitemShut
  {NoStop}%
\bibitem [{\citenamefont {Savasta}\ \emph {et~al.}(2005)\citenamefont
  {Savasta}, \citenamefont {Stefano}, \citenamefont {Savona},\ and\
  \citenamefont {Langbein}}]{Savasta_PRL94_246401_05}%
  \BibitemOpen
  \bibfield  {author} {\bibinfo {author} {\bibfnamefont {S.}~\bibnamefont
  {Savasta}}, \bibinfo {author} {\bibfnamefont {O.~D.}\ \bibnamefont
  {Stefano}}, \bibinfo {author} {\bibfnamefont {V.}~\bibnamefont {Savona}},\
  and\ \bibinfo {author} {\bibfnamefont {W.}~\bibnamefont {Langbein}},\
  }\bibfield  {title} {\bibinfo {title} {Quantum complementarity of microcavity
  polaritons},\ }\href {https://doi.org/10.1103/PhysRevLett.94.246401}
  {\bibfield  {journal} {\bibinfo  {journal} {Phys. Rev. Lett.}\ }\textbf
  {\bibinfo {volume} {94}},\ \bibinfo {pages} {246401} (\bibinfo {year}
  {2005})}\BibitemShut {NoStop}%
\bibitem [{\citenamefont {Romanelli}\ \emph {et~al.}(2007)\citenamefont
  {Romanelli}, \citenamefont {Leyder}, \citenamefont {Karr}, \citenamefont
  {Giacobino},\ and\ \citenamefont {Bramati}}]{Romanelli_PRL98_106401_07}%
  \BibitemOpen
  \bibfield  {author} {\bibinfo {author} {\bibfnamefont {M.}~\bibnamefont
  {Romanelli}}, \bibinfo {author} {\bibfnamefont {C.}~\bibnamefont {Leyder}},
  \bibinfo {author} {\bibfnamefont {J.~P.}\ \bibnamefont {Karr}}, \bibinfo
  {author} {\bibfnamefont {E.}~\bibnamefont {Giacobino}},\ and\ \bibinfo
  {author} {\bibfnamefont {A.}~\bibnamefont {Bramati}},\ }\bibfield  {title}
  {\bibinfo {title} {Four wave mixing oscillation in a semiconductor
  microcavity: Generation of two correlated polariton populations},\ }\href
  {https://doi.org/10.1103/PhysRevLett.98.106401} {\bibfield  {journal}
  {\bibinfo  {journal} {Phys. Rev. Lett.}\ }\textbf {\bibinfo {volume} {98}},\
  \bibinfo {pages} {106401} (\bibinfo {year} {2007})}\BibitemShut {NoStop}%
\bibitem [{\citenamefont {Portolan}\ \emph {et~al.}(2010)\citenamefont
  {Portolan}, \citenamefont {Stefano}, \citenamefont {Savasta},\ and\
  \citenamefont {Savona}}]{Portolan_JPCS210_12033_10}%
  \BibitemOpen
  \bibfield  {author} {\bibinfo {author} {\bibfnamefont {S.}~\bibnamefont
  {Portolan}}, \bibinfo {author} {\bibfnamefont {O.~D.}\ \bibnamefont
  {Stefano}}, \bibinfo {author} {\bibfnamefont {S.}~\bibnamefont {Savasta}},\
  and\ \bibinfo {author} {\bibfnamefont {V.}~\bibnamefont {Savona}},\
  }\bibfield  {title} {\bibinfo {title} {Emergent entanglement of microcavity
  polariton pairs},\ }\href {https://doi.org/10.1088/1742-6596/210/1/012033}
  {\bibfield  {journal} {\bibinfo  {journal} {Journal of Physics: Conference
  Series}\ }\textbf {\bibinfo {volume} {210}},\ \bibinfo {pages} {012033}
  (\bibinfo {year} {2010})}\BibitemShut {NoStop}%
\bibitem [{\citenamefont {Diederichs}\ \emph {et~al.}(2006)\citenamefont
  {Diederichs}, \citenamefont {Tignon}, \citenamefont {Dasbach}, \citenamefont
  {Ciuti}, \citenamefont {Lema{\^i}tre}, \citenamefont {Bloch}, \citenamefont
  {Roussignol1},\ and\ \citenamefont {Delalande}}]{Diederichs_N440_904_06}%
  \BibitemOpen
  \bibfield  {author} {\bibinfo {author} {\bibfnamefont {C.}~\bibnamefont
  {Diederichs}}, \bibinfo {author} {\bibfnamefont {J.}~\bibnamefont {Tignon}},
  \bibinfo {author} {\bibfnamefont {G.}~\bibnamefont {Dasbach}}, \bibinfo
  {author} {\bibfnamefont {C.}~\bibnamefont {Ciuti}}, \bibinfo {author}
  {\bibfnamefont {A.}~\bibnamefont {Lema{\^i}tre}}, \bibinfo {author}
  {\bibfnamefont {J.}~\bibnamefont {Bloch}}, \bibinfo {author} {\bibfnamefont
  {P.}~\bibnamefont {Roussignol1}},\ and\ \bibinfo {author} {\bibfnamefont
  {C.}~\bibnamefont {Delalande}},\ }\bibfield  {title} {\bibinfo {title}
  {Parametric oscillation in vertical triple microcavities},\ }\href
  {https://doi.org/10.1038/nature04602} {\bibfield  {journal} {\bibinfo
  {journal} {Nature}\ }\textbf {\bibinfo {volume} {440}},\ \bibinfo {pages}
  {904} (\bibinfo {year} {2006})}\BibitemShut {NoStop}%
\bibitem [{\citenamefont {Ferrier}\ \emph {et~al.}(2010)\citenamefont
  {Ferrier}, \citenamefont {Pigeon}, \citenamefont {Wertz}, \citenamefont
  {Bamba}, \citenamefont {Senellart}, \citenamefont {Sagnes}, \citenamefont
  {Lema\^{i}tre}, \citenamefont {Ciuti},\ and\ \citenamefont
  {Bloch}}]{Ferrier_APL97_31105_10}%
  \BibitemOpen
  \bibfield  {author} {\bibinfo {author} {\bibfnamefont {L.}~\bibnamefont
  {Ferrier}}, \bibinfo {author} {\bibfnamefont {S.}~\bibnamefont {Pigeon}},
  \bibinfo {author} {\bibfnamefont {E.}~\bibnamefont {Wertz}}, \bibinfo
  {author} {\bibfnamefont {M.}~\bibnamefont {Bamba}}, \bibinfo {author}
  {\bibfnamefont {P.}~\bibnamefont {Senellart}}, \bibinfo {author}
  {\bibfnamefont {I.}~\bibnamefont {Sagnes}}, \bibinfo {author} {\bibfnamefont
  {A.}~\bibnamefont {Lema\^{i}tre}}, \bibinfo {author} {\bibfnamefont
  {C.}~\bibnamefont {Ciuti}},\ and\ \bibinfo {author} {\bibfnamefont
  {J.}~\bibnamefont {Bloch}},\ }\bibfield  {title} {\bibinfo {title} {Polariton
  parametric oscillation in a single micropillar cavity},\ }\href
  {https://doi.org/10.1063/1.3466902} {\bibfield  {journal} {\bibinfo
  {journal} {Appl. Phys. Lett.}\ }\textbf {\bibinfo {volume} {97}},\ \bibinfo
  {eid} {031105} (\bibinfo {year} {2010})}\BibitemShut {NoStop}%
\bibitem [{\citenamefont {Kuznetsov}\ \emph {et~al.}(2019)\citenamefont
  {Kuznetsov}, \citenamefont {Biermann},\ and\ \citenamefont
  {Santos}}]{PVS318}%
  \BibitemOpen
  \bibfield  {author} {\bibinfo {author} {\bibfnamefont {A.~S.}\ \bibnamefont
  {Kuznetsov}}, \bibinfo {author} {\bibfnamefont {K.}~\bibnamefont
  {Biermann}},\ and\ \bibinfo {author} {\bibfnamefont {P.~V.}\ \bibnamefont
  {Santos}},\ }\bibfield  {title} {\bibinfo {title} {Dynamic acousto-optical
  control of confined polariton condensates: From single traps to coupled
  lattices},\ }\href {https://doi.org/10.1103/PhysRevResearch.1.023030}
  {\bibfield  {journal} {\bibinfo  {journal} {Phys. Rev. Research}\ }\textbf
  {\bibinfo {volume} {1}},\ \bibinfo {pages} {023030} (\bibinfo {year}
  {2019})}\BibitemShut {NoStop}%
\bibitem [{\citenamefont {Kuznetsov}\ \emph {et~al.}(2018)\citenamefont
  {Kuznetsov}, \citenamefont {Helgers}, \citenamefont {Biermann},\ and\
  \citenamefont {Santos}}]{PVS312}%
  \BibitemOpen
  \bibfield  {author} {\bibinfo {author} {\bibfnamefont {A.~S.}\ \bibnamefont
  {Kuznetsov}}, \bibinfo {author} {\bibfnamefont {P.~L.~J.}\ \bibnamefont
  {Helgers}}, \bibinfo {author} {\bibfnamefont {K.}~\bibnamefont {Biermann}},\
  and\ \bibinfo {author} {\bibfnamefont {P.~V.}\ \bibnamefont {Santos}},\
  }\bibfield  {title} {\bibinfo {title} {{Quantum confinement of
  exciton-polaritons in structured (Al,Ga)As microcavity}},\ }\href
  {https://doi.org/10.1103/PhysRevB.97.195309} {\bibfield  {journal} {\bibinfo
  {journal} {Phys. Rev. B}\ }\textbf {\bibinfo {volume} {97}},\ \bibinfo
  {pages} {195309} (\bibinfo {year} {2018})}\BibitemShut {NoStop}%
\bibitem [{\citenamefont {Sogawa}\ \emph {et~al.}(2001)\citenamefont {Sogawa},
  \citenamefont {Santos}, \citenamefont {Zhang}, \citenamefont {Eshlaghi},
  \citenamefont {Wieck},\ and\ \citenamefont {Ploog}}]{PVS107}%
  \BibitemOpen
  \bibfield  {author} {\bibinfo {author} {\bibfnamefont {T.}~\bibnamefont
  {Sogawa}}, \bibinfo {author} {\bibfnamefont {P.~V.}\ \bibnamefont {Santos}},
  \bibinfo {author} {\bibfnamefont {S.~K.}\ \bibnamefont {Zhang}}, \bibinfo
  {author} {\bibfnamefont {S.}~\bibnamefont {Eshlaghi}}, \bibinfo {author}
  {\bibfnamefont {A.~D.}\ \bibnamefont {Wieck}},\ and\ \bibinfo {author}
  {\bibfnamefont {K.~H.}\ \bibnamefont {Ploog}},\ }\bibfield  {title} {\bibinfo
  {title} {Dynamic band structure modulation of quantum wells by surface
  acoustic waves},\ }\href@noop {} {\bibfield  {journal} {\bibinfo  {journal}
  {Phys. Rev. B}\ }\textbf {\bibinfo {volume} {63}},\ \bibinfo {pages}
  {121307(R)} (\bibinfo {year} {2001})}\BibitemShut {NoStop}%
\bibitem [{\citenamefont {{de Lima, Jr.}}\ and\ \citenamefont
  {Santos}(2005)}]{PVS156}%
  \BibitemOpen
  \bibfield  {author} {\bibinfo {author} {\bibfnamefont {M.~M.}\ \bibnamefont
  {{de Lima, Jr.}}}\ and\ \bibinfo {author} {\bibfnamefont {P.~V.}\
  \bibnamefont {Santos}},\ }\bibfield  {title} {\bibinfo {title} {Modulation of
  photonic structures by surface acoustic waves},\ }\href
  {https://doi.org/10.1088/0034-4885/68/7/r02} {\bibfield  {journal} {\bibinfo
  {journal} {Rep. Prog. Phys.}\ }\textbf {\bibinfo {volume} {68}},\ \bibinfo
  {pages} {1639} (\bibinfo {year} {2005})}\BibitemShut {NoStop}%
\bibitem [{\citenamefont {Ciuti}\ \emph {et~al.}(2001)\citenamefont {Ciuti},
  \citenamefont {Schwendimann},\ and\ \citenamefont
  {Quattropani}}]{Ciuti_PRB63_41303_01}%
  \BibitemOpen
  \bibfield  {author} {\bibinfo {author} {\bibfnamefont {C.}~\bibnamefont
  {Ciuti}}, \bibinfo {author} {\bibfnamefont {P.}~\bibnamefont
  {Schwendimann}},\ and\ \bibinfo {author} {\bibfnamefont {A.}~\bibnamefont
  {Quattropani}},\ }\bibfield  {title} {\bibinfo {title} {Parametric
  luminescence of microcavity polaritons},\ }\href
  {https://doi.org/10.1103/PhysRevB.63.041303} {\bibfield  {journal} {\bibinfo
  {journal} {Phys. Rev. B}\ }\textbf {\bibinfo {volume} {63}},\ \bibinfo
  {pages} {041303} (\bibinfo {year} {2001})}\BibitemShut {NoStop}%
\bibitem [{\citenamefont {Carusotto}\ and\ \citenamefont
  {Ciuti}(2005)}]{Carusotto_PRB72_125335_05}%
  \BibitemOpen
  \bibfield  {author} {\bibinfo {author} {\bibfnamefont {I.}~\bibnamefont
  {Carusotto}}\ and\ \bibinfo {author} {\bibfnamefont {C.}~\bibnamefont
  {Ciuti}},\ }\bibfield  {title} {\bibinfo {title} {Spontaneous
  microcavity-polariton coherence across the parametric threshold: Quantum
  {Monte Carlo} studies},\ }\href {https://doi.org/10.1103/PhysRevB.72.125335}
  {\bibfield  {journal} {\bibinfo  {journal} {Phys. Rev. B}\ }\textbf {\bibinfo
  {volume} {72}},\ \bibinfo {pages} {125335} (\bibinfo {year}
  {2005})}\BibitemShut {NoStop}%
\bibitem [{\citenamefont {Rudolph}\ \emph {et~al.}(2007)\citenamefont
  {Rudolph}, \citenamefont {Hey},\ and\ \citenamefont {Santos}}]{PVS177}%
  \BibitemOpen
  \bibfield  {author} {\bibinfo {author} {\bibfnamefont {J.}~\bibnamefont
  {Rudolph}}, \bibinfo {author} {\bibfnamefont {R.}~\bibnamefont {Hey}},\ and\
  \bibinfo {author} {\bibfnamefont {P.~V.}\ \bibnamefont {Santos}},\ }\bibfield
   {title} {\bibinfo {title} {Long-range exciton transport by dynamic strain
  fields in a {GaAs} quantum well},\ }\href
  {https://doi.org/10.1103/PhysRevLett.99.047602} {\bibfield  {journal}
  {\bibinfo  {journal} {Phys. Rev. Lett.}\ }\textbf {\bibinfo {volume} {99}},\
  \bibinfo {pages} {047602} (\bibinfo {year} {2007})}\BibitemShut {NoStop}%
\bibitem [{\citenamefont {LaBella}\ \emph {et~al.}(2000)\citenamefont
  {LaBella}, \citenamefont {Bullock}, \citenamefont {Ding}, \citenamefont
  {Emery}, \citenamefont {Harter},\ and\ \citenamefont
  {Thibado}}]{LaBella_JVSTA18_1526_00}%
  \BibitemOpen
  \bibfield  {author} {\bibinfo {author} {\bibfnamefont {V.~P.}\ \bibnamefont
  {LaBella}}, \bibinfo {author} {\bibfnamefont {D.~W.}\ \bibnamefont
  {Bullock}}, \bibinfo {author} {\bibfnamefont {Z.}~\bibnamefont {Ding}},
  \bibinfo {author} {\bibfnamefont {C.}~\bibnamefont {Emery}}, \bibinfo
  {author} {\bibfnamefont {W.~G.}\ \bibnamefont {Harter}},\ and\ \bibinfo
  {author} {\bibfnamefont {P.~M.}\ \bibnamefont {Thibado}},\ }\bibfield
  {title} {\bibinfo {title} {Monte carlo derived diffusion parameters for {Ga}
  on the {GaAs}(001)- (2x4) surface: A molecular beam epitaxy scanning
  tunneling microscopy study},\ }\href {https://doi.org/10.1116/1.582379}
  {\bibfield  {journal} {\bibinfo  {journal} {J. Vac. Sci. Technol., A}\
  }\textbf {\bibinfo {volume} {18}},\ \bibinfo {pages} {1526} (\bibinfo {year}
  {2000})},\ \Eprint {https://arxiv.org/abs/http://dx.doi.org/10.1116/1.582379}
  {http://dx.doi.org/10.1116/1.582379} \BibitemShut {NoStop}%
\bibitem [{\citenamefont {{de Lima, Jr.}}\ \emph {et~al.}(2006)\citenamefont
  {{de Lima, Jr.}}, \citenamefont {van~der Poel}, \citenamefont {Santos},\ and\
  \citenamefont {Hvam}}]{PVS169}%
  \BibitemOpen
  \bibfield  {author} {\bibinfo {author} {\bibfnamefont {M.~M.}\ \bibnamefont
  {{de Lima, Jr.}}}, \bibinfo {author} {\bibfnamefont {M.}~\bibnamefont
  {van~der Poel}}, \bibinfo {author} {\bibfnamefont {P.~V.}\ \bibnamefont
  {Santos}},\ and\ \bibinfo {author} {\bibfnamefont {J.~M.}\ \bibnamefont
  {Hvam}},\ }\bibfield  {title} {\bibinfo {title} {Phonon-induced polariton
  superlattices},\ }\href@noop {} {\bibfield  {journal} {\bibinfo  {journal}
  {Phys. Rev. Lett.}\ }\textbf {\bibinfo {volume} {97}},\ \bibinfo {pages}
  {045501} (\bibinfo {year} {2006})}\BibitemShut {NoStop}%
\bibitem [{\citenamefont {Dennis}\ \emph {et~al.}(2013)\citenamefont {Dennis},
  \citenamefont {Hope},\ and\ \citenamefont {Johnsson}}]{Dennis_CPC184_201_13}%
  \BibitemOpen
  \bibfield  {author} {\bibinfo {author} {\bibfnamefont {G.~R.}\ \bibnamefont
  {Dennis}}, \bibinfo {author} {\bibfnamefont {J.~J.}\ \bibnamefont {Hope}},\
  and\ \bibinfo {author} {\bibfnamefont {M.~T.}\ \bibnamefont {Johnsson}},\
  }\bibfield  {title} {\bibinfo {title} {Xmds2: Fast, scalable simulation of
  coupled stochastic partial differential equations},\ }\href
  {https://doi.org/https://doi.org/10.1016/j.cpc.2012.08.016} {\bibfield
  {journal} {\bibinfo  {journal} {Computer Physics Communications}\ }\textbf
  {\bibinfo {volume} {184}},\ \bibinfo {pages} {201} (\bibinfo {year}
  {2013})}\BibitemShut {NoStop}%
\end{thebibliography}

%

\end{document}


\preprint{SAW-OPO}

\title{Supplementary Material: \\Dynamically  Tuned Arrays of Polariton Parametric Oscillators}


\author{Alexander S. Kuznetsov}
\email{kuznetsov@pdi-berlin.de}
\affiliation{Paul-Drude-Institut f{\"u}r Festk{\"o}rperelektronik, Leibniz-Institut im Forschungsverbund Berlin e.~V., Hausvogteiplatz 5-7, 10117 Berlin, Germany }

\author{Galbadrakh Dagvadorj}
\affiliation{Department of Physics and Astronomy, University College London, Gower Street London WC1E 6BT, United Kingdom }
\affiliation{Department of Physics, University of Warwick, Coventry CV4 7AL, United Kingdom}

\author{Klaus Biermann}
\affiliation{Paul-Drude-Institut f{\"u}r Festk{\"o}rperelektronik, Leibniz-Institut im Forschungsverbund Berlin e.~V., Hausvogteiplatz 5-7, 10117 Berlin, Germany }

\author{Marzena Szymanska}
\affiliation{Department of Physics and Astronomy, University College London, Gower Street London WC1E 6BT, United Kingdom }

\author{Paulo V. Santos}
\email{santos@pdi-berlin.de}
\affiliation{Paul-Drude-Institut f{\"u}r Festk{\"o}rperelektronik, Leibniz-Institut im Forschungsverbund Berlin e.~V., Hausvogteiplatz 5-7, 10117 Berlin, Germany }

\date{\today}

\maketitle



\beginsupplement

This document provides supplementary information to the experiments and data analysis presented in the main text.
We start with studies carried out to determine the potential profile  (Sec.~\ref{SM:Intra-cavity_polariton_traps}) of the intra-cavity traps as well as the energy levels of confined  polaritons in these traps (Sec.~\ref{SM:Energy_levels_of_intra-cavity_traps}). 
The modulation of the confined states by a SAW is analyzed in Sec.~\ref{SM:Acoustic_Modulation}.
The next two sections summarizes results on the dependence of the OPO properties on the frequency of the acoustic excitation  (Sec.~\ref{SM:Frequency}) and on the energy of the pumping laser beam (Sec.~\ref{SM:Pump}).

\section{Structure of the intra-cavity polariton traps}
\label{SM:Intra-cavity_polariton_traps}

\begin{figure}[!tbhp]
    \centering
    \includegraphics[width=\columnwidth, angle=0, clip]{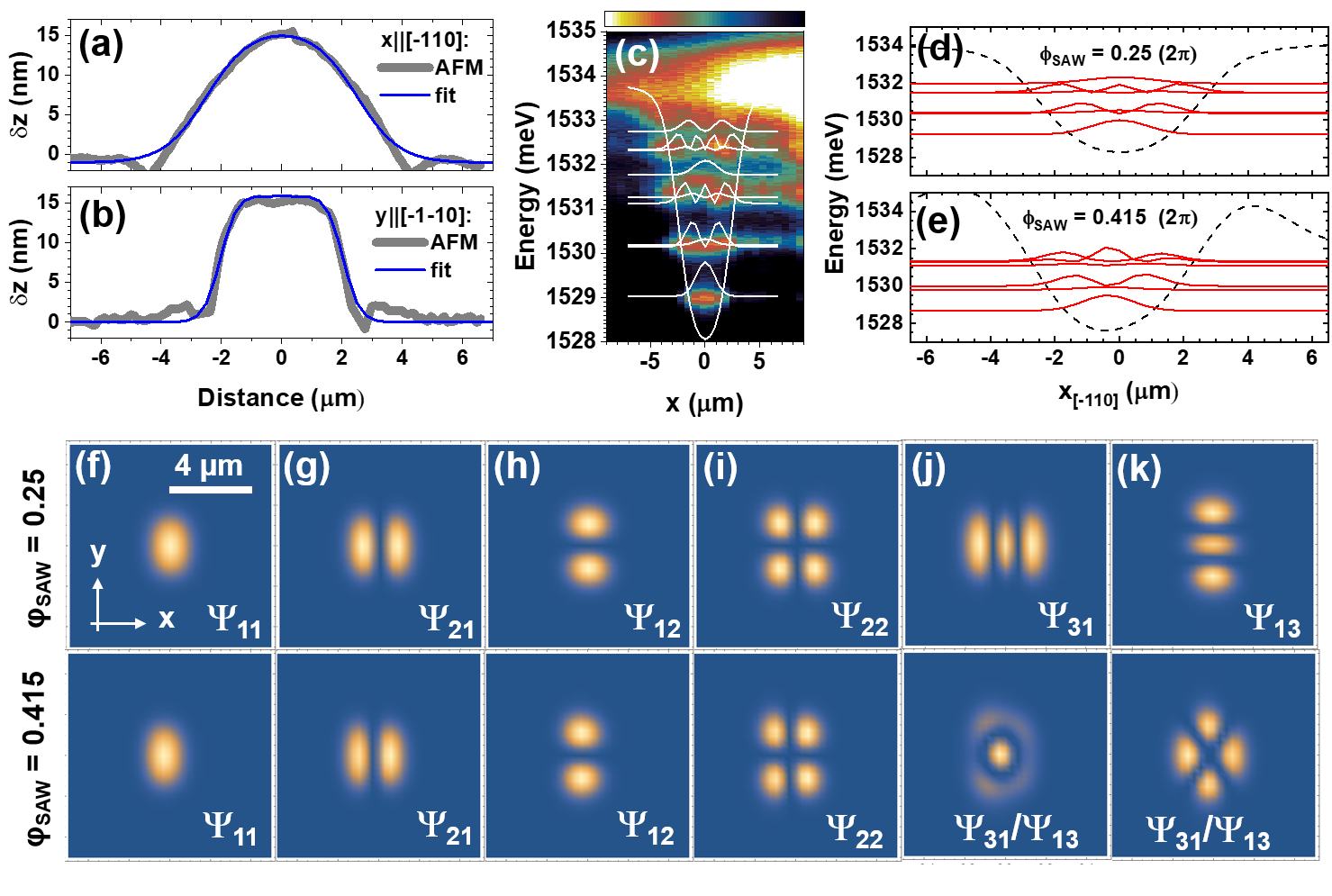}
    \caption{(a)-(b) (thick grey lines) Surface relief measured by scanning an atomic force microscopy (AFM) tip over a surface area on the MC containing an intra-cavity trap with nominal dimensions  of $4\times 4~\mu$m$^2$ along the $x=[\bar110]$ and $y=[\bar1\bar10]$ crystallographic axes, respectively. The thin blue lines are fits with Eq.~{\ref{SM:Eqdz}} yielding the parameters listed in Table~\ref{tab:SMT1}. 
    (c) Comparison of the spatially resolved PL map  with the squared wave functions projections $|\Psi_{m_x m_y}(x,y)|^2$   calculated for the first confined states ($m_x, m_y=1,2,3,4$). (d)-(e) Calculated wave functions for two phases $\phi_\mathrm{SAW}$ of the SAW potential in Eq.~{\ref{EqVSAW}} of the main text. 
    (f)-(k) Comparison between the squared polariton wave functions in the absence (upper panels) and presence of the SAW potential (lower panels) corresponding to panels (d) and (e) respectively. Note that the state for the phase $\phi_{SAW} = 0.415$ in (k) is a superposition of the $\Psi_{31}$ and $\Psi_{13}$ states.
      }
    \label{FigSM1}    
\end{figure}

The MBE overgrowth of the structured microcavity (MC) spacer results in a relief of the sample surface, which reflects the structural dimensions of the polariton intra-cavity traps embedded in the spacer \cite{PVS312}. The thick lines in Figs.~\ref{FigSM1}(a) and \ref{FigSM1}(b) show the surface relief of the final MC induced by an intra-cavity trap measured by scanning an atomic force microscopy (AFM) along the $x=[\bar110]$ and $y=[\bar1\bar10]$ crystallographic axes, respectively. The mesa with a nominal height of 12 and lateral dimensions of $4\times 4~\mu$m$^2$ etched in the MC spacer gives rise to a surface relief with a height of $\delta z_0 = 15$~nm after the MBE overgrowth. While the height profiles  $\delta z(r_i)$ ($r_i=x, y$) along the $y$ axis closely reproduce the dimension of the mesa etched in the MC spacer,  the profile shapes along $x$ are significantly smoother. This behavior is attributed to the anisotropic dynamics of the MBE overgrowth process \cite{LaBella_JVSTA18_1526_00,PVS312}.
Following the procedure delineated in Ref.~\cite{PVS312}, we obtained the spatial dependence of the changes in trap height $\delta z(r_i)$   by fitting the AFM profiles along the two directions to the following expression:

\begin{equation}
\label{SM:Eqdz}
\delta z (r_i) =\delta z_0 
\left[ 	\textit{Erfc}\left(\frac{r_i -  w_i/2}{ \sqrt{2} w_i} \right) + 
  		\textit{Erfc}\left(\frac{r_i + w_i/2}{ \sqrt{2} w_i} \right)\right].
\end{equation} 

\noindent Here, $\textit{Erfc}(r)$ is the complementary error function and $w_i$ the effective width of the lateral interface between non-etched and etched regions. The thin blue lines superimposed on the AFM curves of Figs.~\ref{FigSM1}(a) and \ref{FigSM1}(b) display fits of Eq.~(\ref{SM:Eqdz}) to the experimental data, which yield the structural parameters for the intra-cavity trap summarized in Table~\ref{tab:SMT1}. \dAK{\aPVS{{\bf AK: please complete Table~\ref{tab:SMT1}} by including the missing values! }}

\begin{table}[!htbp]
\label{tab:SMT1}
\caption{Structural parameters and simulations parameters for intra-cavity traps.}
  \begin{threeparttable}[t]
  \centering
       \begin{tabular}{ l c }
    \toprule
     Parameter     & Value\\
    Lateral mesa dimensions, $l_x$ and $l_y$ 	            & $6.1/4.5~\mu$m\\
    Lateral interface widths, $w_x$ and $w_y$        	& $1.3/0.5~\mu$m\\
      Potential barrier, $E_b$                          		& $6$~meV \\
      Polariton effective mass, $m_p$  						& $5\times 10^{-5}~m_0$\tnote{1} \\
      Rabi energy, $\hbar\Omega_R$                          & 7.2~meV \\
    \toprule
  \end{tabular}
     \begin{tablenotes}
     \item[1] $m_0$ is free electron mass.
   \end{tablenotes}
    \end{threeparttable}%
\end{table}%

\section{Energy levels of intra-cavity traps}
\label{SM:Energy_levels_of_intra-cavity_traps}

The change $\delta z_0$ in the MC spacer thickness results in an energy difference (barrier) of approx. $E_b=6$~meV between the lower polariton states in the non-etched and etched regions of the MC. The energy levels and wave functions for  polaritons confined in intra-cavity traps were determined  by numerically solving the two-dimensional Schrödinger equation for a polariton confinement potential with depth $E_b$ and the  lateral shape  for $\delta z$ given by Eq.~(\ref{SM:Eqdz}) and Table~(\ref{tab:SMT1}). As described in details in Ref.~\cite{PVS312}, the calculations were carried out in Fourier space for a periodic super-cell containing the intra-cavity trap surrounded by  potential barriers. The latter were chosen to be wide enough to avoid interactions between neighboring traps.

The solid lines in Fig.~\ref{FigSM1}(c) displays the squared wave functions $|\Psi_{m_x m_y}(0,y)|^2$ calculated for the first confined states ($m_x, m_y=1,2,3$)  of the the intra-cavity trap. The dashed line shows the polariton confinement potential along the $x$ direction.  The calculations reproduce well the measured energy of the confined states [cf.~Fig.~\ref{FigSM1}(b)] and yield spatial extent for their wave functions very similar to the ones obtained from the PL maps.

\section{Acoustic modulation}
\label{SM:Acoustic_Modulation}

Different mechanisms mediating the modulation of polariton levels by SAW are reviewed in Ref.~\onlinecite{PVS169}. Here, we will make the approximation that the standing SAW field along $x$ induces a modulation of the polariton levels given by Eq.~(\ref{EqVSAW}) of the main text with the field anti-nodes  coinciding with the center of the trap at  $x=0$. Due to relatively low SAW frequency (384~MHz), one can safely assume that the polariton states adiabatically follow the potential variations. The impact of the standing SAW field on the polariton levels was studied by adding to the confinement potential the acoustic potential  given by Eq.~(\ref{EqVSAW}). 
Figures~\ref{FigSM1}(d) and \ref{FigSM1}(e) compare profiles for the wave functions  $|\Psi_{m_x m_y}(x,0)|^2$ for phases  $\phi_\mathrm{SAW}=0.25\pi$~rad, when the standing SAW potential vanishes, with the one for $\phi_\mathrm{SAW}=0.415\pi$~rad and a SAW potential amplitude $V_\mathrm{SAW,0}=2$~meV. The latter yields the energy matching for OPO excitation.  \dAK{\aPVS{{\bf AK: the results in the figure are not consistent! please check!}}}

\subsection{Acoustically induced energy tunning}
\label{Acoustically_induced_energy_tunning}

The symmetry and time evolution of the polariton levels displayed in Fig.~\ref{FigSM1} (as well as in Fig.~\ref{FigOPO}(e) of the main text) can be understood by treating the acoustic potential of Eq.~(\ref{EqVSAW}) as a perturbation of the confined states  of  the rectangular potential given by Eq.~(\ref{EqWF}). For small ratios $\ell_x/\lSAW$ and $\ell_y/\lSAW$, the energy changes $\Delta E_{m_xm_y}$ of the confined states $\Psi_\mathrm{m_x m_y}$ can then we written as:

\begin{equation}
\Delta E_\mathrm{m_x m_y} = 
\langle\Psi_\mathrm{m_xm_y}|V_\mathrm{SAW}|\Psi_\mathrm{m_xm_y} \rangle 
\approx V_\mathrm{SAW,0}   \left(  	\frac{\pi^2}{6} - \frac{1}{m_x^2} \right) 
\left(\frac{l_x}{\lSAW}\right)^2 \cos{(\pSAW)} .
\label{EqSMSIDESAW}
\end{equation}

The approximation in Eq.~(\ref{EqSMSIDESAW}) applies up to second order terms in the ratios $\ell_x/\lSAW$. 

The acoustically induced shift given by Eq.~(\ref{EqSMSIDESAW}) enables tuning the energy levels of the trap by adjusting the SAW amplitude $V_\mathrm{SAW,0}$ to satisfy the OPO energy matching requirement 

\begin{equation}
E_\mathrm{idler} - E_\mathrm{pump} = E_\mathrm{pump}-E_\mathrm{signal}.
\label{EqSIem}
\end{equation}

\noindent Here the subscripts denote the OPO states. By properly selecting $V_\mathrm{SAW,0}$ and $\lSAW$, it is possible to satisfy the OPO energy matching  criterion over a wide range of trap geometries and OPO configurations. Two examples are described below:

\begin{itemize}
	\item States $O_1=\{ \Psi_\mathrm{signal},  \Psi_\mathrm{pump},  \Psi_\mathrm{idler} \}=\{\Psi_{11}, \Psi_{12}\text{ or }\Psi_{21}, \Psi_{22}\}$ in a rectangular trap: these states are energetically equidistant for a perfectly square trap with square size $\ell$, but not in a rectangular with sides $\ell_x\neq\ell_y$. If $l_x=\ell$ and $\ell_y=(1+r_\ell) \ell$, it can be shown using Eq.~(\ref{EqSMSIDESAW}) that the levels become equidistant if:

\begin{equation}
V_\mathrm{SAW,0} \cos{(\pSAW)} = 4 r_\ell \left(\frac{\lSAW}{\ell}\right)^2 E_{11}, 
 \enskip\text{with } \enskip E_{11} = (1-r_\ell) \frac{\pi ^2 \hbar ^2}{m_p \ell^2 }.
\label{EqSIT1}
\end{equation}

\noindent Here, $m_p$ is the polariton mass and $E_{11}$  the zero-motion energy in the trap (corresponding to the quantum shift of the lowest confined level). We note, however, that this set of states miss the symmetry requirements for non-linear coupling conditions required for OPO excitation, as discussed in Sec.~\ref{Symmetry of the OPO states} of the main text.

	\item States $O_2=\{ \Psi_\mathrm{signal}, \Psi_\mathrm{pump}, \Psi_\mathrm{idler} \}=\{\Psi_{11}, \Psi_{21}, \Psi_{13}\}$ in a square trap: if the trap size is  $\ell$, the states becomes equidistant in energy by selecting

\begin{equation}
V_\mathrm{SAW,0}\cos{(\pSAW)} = \frac{2}{3} \left(\frac{\lSAW}{\ell}\right)^2 E_{11}, 
 \enskip \text{with }  \enskip E_{11} = \frac{\pi ^2 \hbar ^2}{m_p \ell^2}.
\label{EqSIT2}
\end{equation}
	
\end{itemize}

\section{Radio-frequency dependence of OPO excitation}
\label{SM:Frequency}

\begin{figure*}[tbhp]
    \centering
    \includegraphics[width=.75\columnwidth, angle=0, clip]{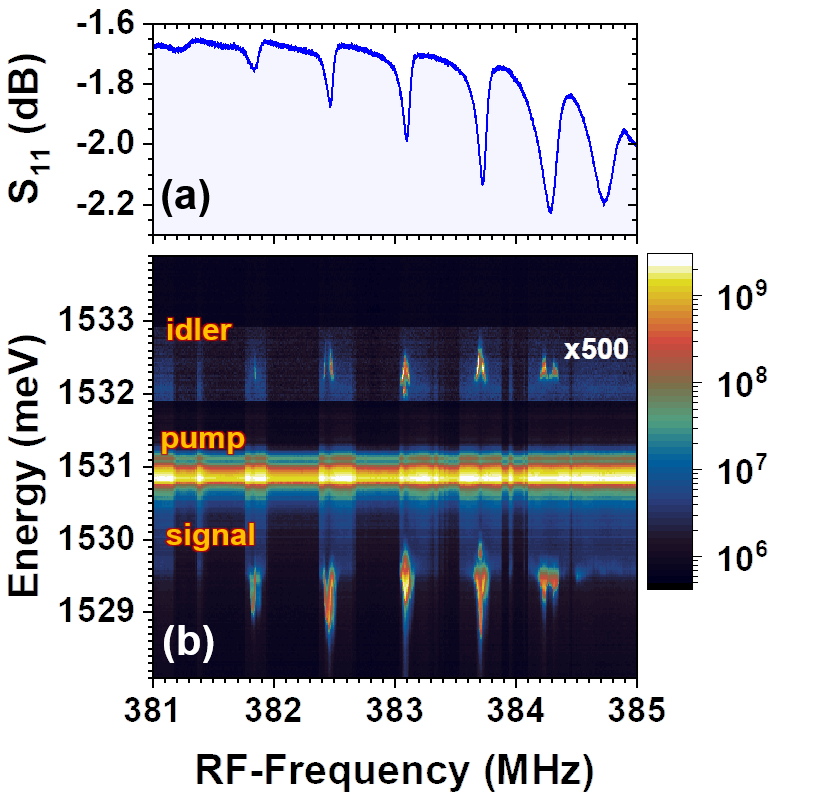}
    \caption{(a) Radio-frequency (rf) power reflection coefficient $|s_{11}|^ 2$ for the delay line in Fig.~\ref{FigStr}. (b) rf dependence of the PL from a $4\times 4~\mu$m$^2$ intracavity traps. Note that OPOs are only excited at the resonance frequencies of the acoustic resonator, which correspond to the dips of the spectrum in (a).}
    \label{FigSM2}
\end{figure*}

 The delay line in Fig.~\ref{FigStr}(c) forms an acoustic resonator due to the acoustic reflections at the single-finger IDTs, which act as acoustic Bragg reflectors. The properties of the resonator were accessed by measuring the radio-frequency (rf) scattering parameters using a vector network analyser. Figure~\ref{FigSM2}(a) displays the spectral dependence of the $|s_{11}|^2$ parameter, which corresponds to the ratio between the reflected and incident rf-power applied to one of the IDTs. The sharp dips in the spectrum correspond to the reduction of the reflected rf power due to the excitation of acoustic modes of the resonator. From the width of these dips one can extract the quality factor of the resonator. The sharpest of the dips at $\fSAW=383.7$~MHz has the quality factor of $Q=4700$.

The map of Fig.~\ref{FigSM2} show in a color scale the spectral dependence of the PL spectrum of the trap on the nominal rf frequency applied to the delay line for a fixed rf level. Note the OPO signal and idler levels only appear at the resonance frequencies of the acoustic resonator, thus unambiguously proving that OPO triggering is due to the acoustic field. In addition, the figure also shows that the energy separation between the pump and signal, while remaining equal the one between idler and pump, changes with frequency following  the changes in amplitude of the SAW strain field. 

\section{OPO dependence on pumping energy}
\label{SM:Pump}

\begin{figure}[t!bhp]
    \centering
    \includegraphics[width=.9\columnwidth, angle=0, clip]{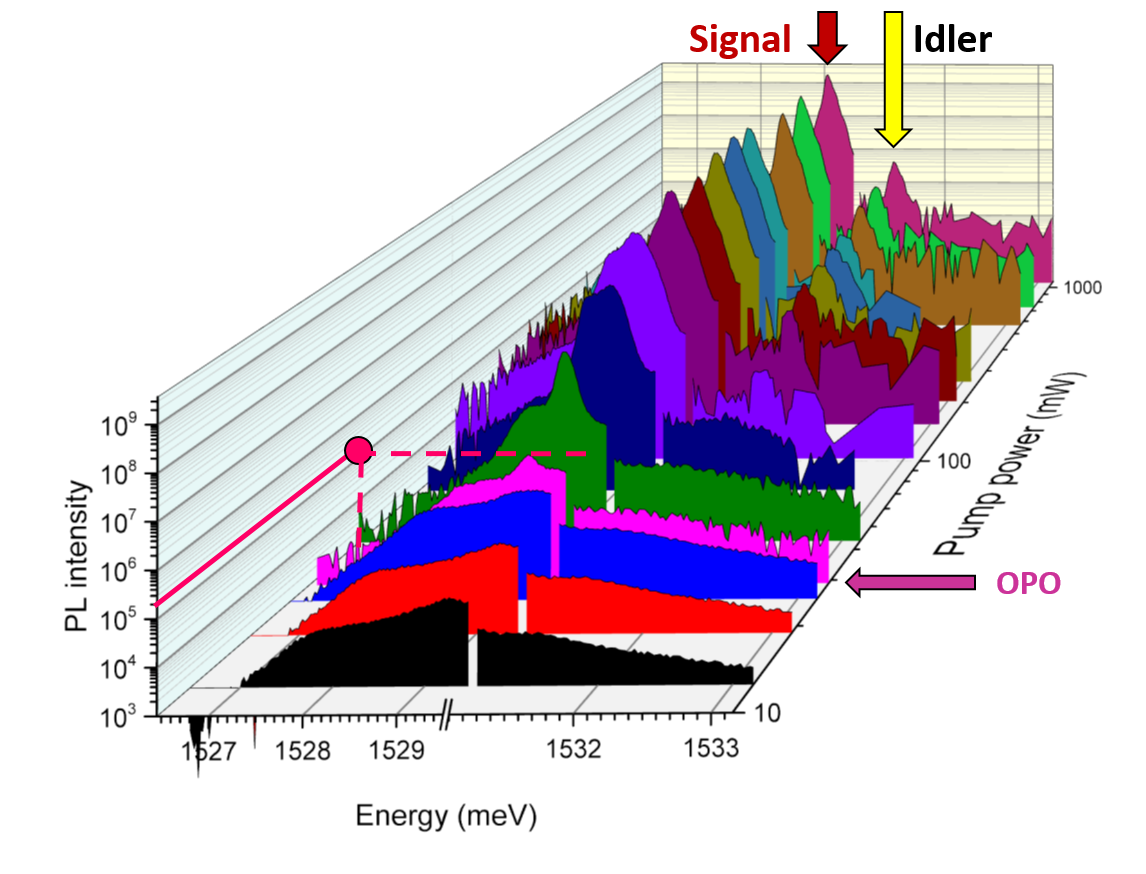}
    \caption{Dependence of the photoluminescence spectrum of a $4\times 4~\mu$m$^2$  intra-cavity trap for a fixed acoustic excitation on the optical pump level. Only the signal and ideler energy ranges are shown. 
    \label{FigPumpPower}}
\end{figure}

The dependence of the OPO properties on the pump intensity is summarized in Fig.~\ref{FigPumpPower}. In this series of plots, we monitored the PL emitted by a $4\times 4~\mu$m$^2$ intra-cavity traps under a fixed amplitude of the SAW field while increasing the optical excitation power from 10 to 1000~mW (focused into a spot of approx. $50~\mu$m in diameter). The OPO triggers at approx. 30~mW (corresponding to a flux of 1.5~kW/cm$^2$). The intensity of the signal and idler states increases with excitation power up to approx. 100~mW and saturates for higher powers. The saturation is attributed to the saturation of pump injection as the pump state blue-shifts due to polariton-polariton interactions. 

The observation of OPO over a relatively wide range of acoustic  (cf.~Fig.~\ref{FigSM2}) and optical  (cf.~Fig.~\ref{FigPumpPower}) excitation conditions indicates that once triggered, non-linearities induced by the polariton interactions permit the OPO over a wide range of excitation conditions.

%
%
\section{Numerical Simulation of the acoustically driven OPO}
\label{Numerical Simulation of the acoustically driven OPO}

We simulate the polariton dynamics by solving Eq.~\ref{EqGPE} in the main text with the XMDS2 software framework \cite{Dennis_CPC184_201_13} using an adaptive step-size fourth and fifth order embedded Runge-Kutta algorithm. In order to allow for \rPVS{the OPO regime, we have used the exact LP dispersion rather than approximating it by a quadratic form, which is only accurate for the low energy polariton modes and  suitable only for non-resonant  excitations}{ OPO we have used whole LP dispersion not approximated by quadratic dispersion which is accurate for only low modes and suitible for non-resonance excitation}. The LP dispersion and external coherent pump are given by:

\begin{equation}
\omega_\mathrm{LP}\left(-i\nabla\right)=\omega_\mathrm{X}+\frac{1}{2}\left[-\frac{\nabla^2}{2m_\mathrm{C}}+\delta-\sqrt{\left(-\frac{\nabla^2}{2m_\mathrm{C}}+\delta\right)^2+\Omega_\mathrm{R}^2}\right],
\label{EqLPD}
\end{equation}
and
\begin{equation}
  F_\mathrm{p}(\textbf{r},t) = f_\mathrm{p} e^{i (\vect{k}_p \cdot \vect{r} - \omega_p t)} \; .
\end{equation}

We choose the system parameters to be close to current experiments:
\begin{itemize}
	\item[$m_C$] $=2.3\times 10^{-5}m_e$: the effective mass of the photons due to the microcavity,
	\item[$\kappa_\mathrm{LP}$] $=0.1$~meV: the decay rate of the lower-polaritons,
	\item[$g_\mathrm{LP}$] $=0.005$~meV$\mu$m$^2$: the polariton-polariton interaction strength,
	\item[$\Omega_R$] $=7.6$~meV: the Rabi splitting and
	\item[$\delta$] $=\omega_C(0)-\omega_X(0)=-5.4$~meV:  the \aPVS{exciton-photon} detuning.
\end{itemize}

The polariton states in the intra-traps, displayed in Figs.~\ref{SimUCL}(a) and \ref{SimUCL}(f) of the main text, were calculated by setting the initial pump power $f_\mathrm{p}$ to zero and the polariton decay rate $\kappa_\mathrm{LP}$ to a very small negative value with a Gaussian noise initial condition. To determine the numerical pump state $\Psi_{21}$ as well as all other \aPVS{states} the final results are averaged over 100 random initial conditions. In order to find the OPO energy matching condition we have taken 100 different values \rPVS{for the SAW phase within a single SAW cycle and averaged,  for each phase point, over 100 initial noise conditions}{of phase across a single SAW cycle and for each phase point we again averaged over 100 initial noises}. 

\IfFileExists{literature.bib}
	{ \def\litdir{.} }
{ \IfFileExists{x:/sawoptik_databases/jabref/literature.bib}  
	{  \def\litdir{x:/sawoptik_databases/jabref} }
      {   \def\litdir{c:/myfiles/jabref}     }
}

\bibliography{\litdir/literature,\litdir/mypapers}